\documentclass[a4paper,fleqn]{cas-dc}

\usepackage[numbers,compress]{natbib}
\graphicspath{{./}{./figs/}}

\def\tsc#1{\csdef{#1}{\textsc{\lowercase{#1}}\xspace}}
\tsc{WGM}
\tsc{QE}
\tsc{EP}
\tsc{PMS}
\tsc{BEC}
\tsc{DE}

\begin{document}
\let\WriteBookmarks\relax
\def\floatpagepagefraction{1}
\def\textpagefraction{.001}
\shorttitle{$\rm S^*(E)$ measurement of $\rm {}^{12}C({}^{12}C,\alpha){}^{20}Ne$ at astrophysical energies via THM ... }
\shortauthors{C. Li et~al.}

\title [mode = title]{$\rm S^*(E)$ measurement of the $\rm {}^{12}C({}^{12}C,\alpha){}^{20}Ne$ reaction at astrophysical energies via the Trojan horse method with $\rm ^{16}O$ quasi-free breakup}                      
\tnotemark[1]

\tnotetext[1]{This document is the results of the research project funded by the National Natural Science Foundation of China (12075031, 12275360) and Natural Science Foundation of Beijing Municipality (1222022).}


\author[1]{Chengbo Li}[
                        orcid=0000-0002-0130-2233]
\cormark[1]
\ead{lichengbo@bjast.ac.cn; lichengbo2008@163.com}

\affiliation[1]{organization={Institute of Radiation Technology, Beijing Academy of Science and Technology},
                city={Beijing},
                postcode={100875}, 
                country={China}}

\author[2]{Huiming Jia}
\cormark[2]
\ead{jiahm@cnncmail.cn}

\affiliation[2]{organization={Department of Nuclear Physics, China Institute of Atomic Energy},
                city={Beijing},
                postcode={102413}, 
                country={China}}

\author[3]{Qungang Wen}[%
   ]
\cormark[3]
\ead{qungang@ahu.edu.cn}

\affiliation[3]{organization={School of Physics and Materials Science, Anhui University},
                city={Hefei},
                postcode={230601}, 
                country={China}}

\author[2]{Chengjian Lin}
\author[2]{Lei Yang}
\author[2]{Feng Yang}
\author[2]{Nanru Ma}
\author[2]{Peiwei Wen}
\author[2]{Tianpeng Luo}
\author[2]{Chang Chang}
\author[1]{Xuepeng Sun}
\author[3]{Xuejian Wang}

\cortext[cor1]{Principal corresponding author}
\cortext[cor2]{Corresponding author}
\cortext[cor3]{Corresponding author}



\begin{abstract}
The $\rm {}^{12}C(^{12}C,\alpha)^{20}Ne$ reaction at astrophysical energies is crucial for understanding the carbon burning process in massive star and explosive astrophysical scenarios like Type Ia supernovae and X-ray bursts. 
However, directly measuring or simply extrapolating its $\rm S^*(E)$ factor is extremely challenging  due to Coulomb suppression and potential complex resonance structures near the Gamow window ($\rm E_G=1.5\pm 0.3$ MeV).
The Trojan horse method (THM) can circumvent the Coulomb barrier, providing data within the Gamow window without extrapolation. 
Strong resonances near 1.5 MeV were previously reported by Tumino et al. (2018) using THM with $\rm ^{14}N= (^{12}C \oplus d)$, a result that generated significant interest and debate, underscoring the need for further experimental verification. 
In this work, we selected $\rm {}^{16}O=({}^{12}C \oplus \alpha)$ as the Trojan-horse nucleus due to its lower binding energy, which favors quasi-free reactions. We performed an indirect measurement of 
$\rm {}^{12}C(^{16}O,\alpha\alpha){}^{20}Ne$ at the HI-13 Tandem Accelerator at CIAE. Employing a copper beam-stopper foil, we measured, for the first time in a THM experiment, the spectator $\alpha$-particle within a small angular range around 0°, where the quasi-free mechanism predicts its highest concentration. 
The $\rm S^*(E)$ factor of $\rm {}^{12}C(^{12}C,\alpha)^{20}Ne$ in the astrophysical energy region was extracted from the measured three-body reaction using THM based on the distorted-wave Born approximation (DWBA). 
Our results confirm the existence of resonances within the Gamow window around 1.5 MeV in both the $\alpha_0$ and $\alpha_1$ channels. 
Without considering the details of the resonance structures, the overall trend of our results is qualitatively in reasonable agreement with the THM-Tumino2018 and TTIK2025 data, but differs significantly from the trend of the Modified-THM-Muk2019 data. 
We observe no evidence for hindrance effect in our results. 
\end{abstract}

\begin{graphicalabstract}
\includegraphics[width=0.5\columnwidth]{THMCC-Oca.pdf}
\end{graphicalabstract}

\begin{highlights}
\item This study successfully utilized $\rm {}^{16}O=({}^{12}C \oplus \alpha)$ as the Trojan horse nucleus,  capitalizing on its low binding energy ($e_a$ = 7.16 MeV).

\item{}
Beam-stopper foil enabled first measurement of spectator particles near 0°, where quasi-free mechanism predicts peak yield.

\item{}
Indirect measurement revealed resonance features near 1.5 MeV, supporting conclusions of earlier indirect work (THM2018).

\item{}
Overall trends agree with THM-Tumino2018 and TTIK2025 data, but differ significantly from Modified-THM-Muk2019 results.

\item We observe no evidence for hindrance in the energy range.

\end{highlights}

\begin{keywords}
Carbon fusion reaction \sep Gamow energy \sep  $\rm S^*(E)$-factor \sep  Trojan horse method \sep Stellar nucleosynthesis
\end{keywords}

\maketitle

\section{\label{sec:intro} Introduction}

The carbon burning reaction $\rm ^{12}C+{}^{12}C$ plays a key role in the late evolution of the massive stars \cite{RMP2014} and in explosive astrophysical environments such as Type Ia supernovae and X-ray bursts \cite{RMP2002, Ast2001}. 
The evolution and nucleosynthesis processes in stars, from medium mass (8-10 $\rm M_{\odot}$) to those over 10 $\rm M_{\odot}$, are strongly influenced by the carbon fusion process. 
The minimum stellar mass required to ignite carbon burning is closely tied to the accurately measured cross section of the $\rm ^{12}C+{}^{12}C$ reaction within the Gamow energy region \cite{RMP2014}. 
Type Ia supernovae explosions result from the thermonuclear disruption of a carbon-oxygen white dwarf in a binary system upon reaching the Chandrasekhar mass limit. During this process, the temperature and density of the material increase dramatically, making the accurate carbon burning reaction rate a key input for modeling the ignition \cite{RMP2002}
The carbon fusion reaction is also believed to trigger X-ray superbursts on the surface of neutron stars following the rp-process \cite{Ast2001}. 
The $\rm ^{12}C+{}^{12}C$ system exhibits a quasi-molecular structure, where the cross section can be significantly enhanced by the presence of quasi-molecular resonances \cite{Spi2007}. Despite numerous attempts, no single theoretical model can satisfactorily explain all the observed details.

The carbon burning reaction in star proceeds through highly-excited states of the compound nucleus $\rm {}^{24}Mg^{*}$ , decaying primarily via the following channels:
$$ 
\begin{matrix}
  & \rm ^{12}C({}^{12}C,\alpha){}^{20}Ne & \rm (Q= 4.62 MeV)\\
  & \rm ^{12}C({}^{12}C,p){}^{23}Na & \rm (Q= 2.24 MeV)\\
  & \rm ^{12}C({}^{12}C,n){}^{23}Mg & \rm (Q= -2.60 MeV)\\
  & \rm ^{12}C({}^{12}C,\gamma){}^{24}Mg  & \rm (Q= 13.93 MeV)\\
  & \rm ^{12}C({}^{12}C,2\alpha){}^{16}O  & \rm (Q= -0.113 MeV) 
\end{matrix}
$$

At energies below 2.6 MeV, the reaction most probable proceed via the $\alpha$ (mainly via $\alpha_0$ and $\alpha_1$ ) and $p$ (mainly via $p_0$ and $p_1$ ) channels \cite{Lyj2020}.
The branching ratios between the $\alpha$ and $p$ channels significantly impact the final elemental abundances.

The temperature for typical stellar carbon burning is around 0.8 GK, corresponding to a center-of-mass energy of $\rm E_G=1.5 \pm 0.3$ MeV, known as the Gamow window.
For an extended temperature range of 0.6 to 1.2 GK, the corresponding energy spans from 1 to 3 MeV. However, the Coulomb barrier of the $\rm ^{12}C+{}^{12}C$  system is approximately 7.8 MeV ($V_c=Z_1 Z_2 e^2/4\pi \varepsilon_0 r_0 (A_1^{1/3}+A_2^{1/3})$, with $r_0$ = 1.45 fm), far exceeding the Gamow energy and making direct measurements at low energies exceedingly difficult. 

\begin{table*}
\begin{center}
    \caption{\label{tab:SummaryCC} Summary of experimental measurements of the $\rm ^{12}C+{}^{12}C$ reaction near astrophysical energies}
        \begin{tabular*}{170mm}{l@{\extracolsep{\fill}}ll}
    \toprule
    Measurement & Method & Energy range $\rm E_{cm}$(MeV)\\
    \midrule
    1969-Patterson\cite{Pat1969}	& Direct measurement of particle: $p_i$,$\alpha_i$	& 3.23-8.75 \\
    1973-Mazarakis\cite{Maz1973}	& Direct measurement of particle: $p_i$,$\alpha_i$	& 2.55-5.01 \\
    1977-High\&Cujec\cite{Hig1977}	& Direct measurement of $\gamma$	& 2.46-5.88 \\
    1980-Kettner\cite{Ket1980}	& Direct measurement of $\gamma$	& 2.45-6.15 \\
    1981-Becker\cite{Bec1981}	& Direct measurement of particle: $p_i$,$\alpha_i$	& 2.8-6.3 \\
    1982-Dasmahapatra\cite{Das1982}	& Direct measurement of $\gamma$	& 4.2-7.0 \\
    2006-Aguilera\cite{Agu2006}	& Direct measurement of $\gamma$	& 4.42-6.48 \\
    2006-Barron-Palos\cite{Bar2006}	& Direct measurement of $\gamma$	& 2.25-6.01 \\
    2007-Spillane\cite{Spi2007}	& Direct measurement of $\gamma$	& 2.10-4.75 \\
    2018-Zickefoose\cite{Zic2018}	& Direct measurement of particle: $p_i$,$\alpha_i$	& 2.1-4.0 \\
    2018-Jiang\cite{Jcl2018}	& Coincidence measurement of $\gamma$ + particle $p_1$,$\alpha_1$	& 2.68-4.93 \\
    2020-Fruet\cite{Fru2020}	& Coincidence measurement of $\gamma$ + particle $p_1$,$\alpha_1$	& 2.16,2.54-3.77,4.75-5.35 \\
    2020-Tan\cite{Tan2020}	& Coincidence measurement of $\gamma$ + particle $p_1$,$\alpha_1$	& 2.2,2.65-3.0,4.1-5.0 \\
    2024-Tan\cite{tan2024}	& Coincidence measurement of $\gamma$ + particle $p_1$,$\alpha_1$	& 2.65-5.0 \\
    2025-Nippert\cite{nip2025}	& Coincidence measurement of $\gamma$ + particle $p_1$,$\alpha_1$	& 2.3-3.4 \\
    2018-Tumino\cite{Tum2018}	& Indirect, Trojan horse method: particle $p$,$\alpha$	& 0.8-2.7\\
    2025-Nan\cite{wyb2025} & Indirect, Thick-target inverse kinematics method: $\gamma$ + particle $p$,$\alpha$ & 0.5-2.3\\
    \bottomrule
      \end{tabular*}
\end{center}
\end{table*}

Research on the $\rm ^{12}C+{}^{12}C$ reaction began in the 1960s \cite{Bro1960, Alm1960}.  It was discovered that the excitation functions for elastic scattering and nuclear reactions in this system exhibited unexpected resonance structures, spurring extensive experimental \cite{Spi2007, Pat1969, Maz1973,  Hig1977, Ket1980, Bec1981, Das1982, Agu2006, Bar2006, Zic2018, Jcl2018, Fru2020, Tan2020, tan2024, nip2025} and theoretical\cite{CF1988, Hin2007, Cop2009, AMD2021, DIM2024} research to understand the underlying mechanism.  
Significant efforts have been made to measure the cross section of $\rm ^{12}C+{}^{12}C$ near astrophysical energies (Table {\ref{tab:SummaryCC}}), employing methods such as charged-particle spectroscopy\cite{Pat1969, Maz1973, Bec1981, Zic2018}, $\gamma$-ray spectrometry \cite{Spi2007, Hig1977, Ket1980, Das1982, Agu2006, Bar2006}, and coincidence measurements of charged particles and $\gamma$-ray to reduce background \cite{Jcl2018, Fru2020, Tan2020, tan2024, nip2025}. However, $\gamma$-ray spectrometry cannot study the $p_0$ and $\alpha_0$ channels which decay to the ground state of $\rm {}^{23}Na$ and $\rm {}^{20}Ne$ without $\gamma$-ray emission. 

Due to the Coulomb barrier, direct measurements of the carbon burning reaction become increasingly difficult at lower energies. The current lower limit for direct data is only about 2.1 MeV.  Spillane \cite{Spi2007}  reported a strong resonance near $\rm E_{cm}$=2.14 MeV, on the high-energy tail of the Gamow peak. If confirmed, this resonance would increase the non-resonant reaction rate for the $\alpha$ decay channel by a factor of five for stars near T=0.8 GK.

Extrapolating from existing direct data to the ultra-low energy region is problematic due to potential low-energy resonances and the disparate trends predicted by different theoretical models \cite{CF1988, Hin2007, Cop2009, AMD2021, DIM2024}. 
The standard estimate by Caughlan and Fowler (CF88) \cite{CF1988} assumes a constant S-factor and is widely used in astrophysical simulations. 
In contrast, Jiang et al. \cite{Hin2007} proposed a hindrance model predicting a reduction in the S-factor at low energies.
Conversely, Cooper et al. \cite{Cop2009} suggested the possibility of a resonance near 1.5 MeV within the Gamow window. 
The experimental confirmation of such resonance structures would significantly impact the currently adopted carbon fusion reaction rates. Recently, Taniguchi et al. predicted low-energy resonances around 1 MeV and 1.5 MeV using antisymmetrized molecular dynamics (AMD) \cite{AMD2021} and various nuclear energy density functionals \cite{DIM2024}.

In 2018, Tumino et al. \cite{Tum2018} reported the first indirect observation of strong resonances near 1.5 MeV within the Gamow window using the Trojan horse method. These resonances implied a reaction rate increase of more than 25 times the standard value at temperatures around 0.5 GK. However, this result has been met with debate \cite{Muk2018,Tum2018b,Muk2019,Muk2022,Txd2019,Znt2020,Bec2020}.
Mukhamedzhanov et al. \cite{Muk2018,Muk2019,Muk2022} questioned whether the plane-wave impulse approximation (PWIA) used in THM could introduce significant deviations for the $\rm ^{12}C+{}^{12}C$ reaction, advocating for corrections due to Coulomb and nuclear final-state interference. 
The rising trend of the $\rm S^*(E)$ factor towards low energies reported by Tumino et al. exceeds the upper limit set by the CC-M3Y+Rep theory \cite{Bec2020} and constraints from the $\rm ^{12}C+{}^{13}C$ reaction \cite{Znt2020}, leading them to propose a modified $\rm S^*(E)$ factor to reduce it \cite{Muk2019}. 
Tumino et al. countered that this modified $\rm S^*(E)$ factor increased incorrectly towards higher energies, inconsistent with direct measurement data \cite{Tum2018b}. 
It should be noted that the rising trend of the $\rm S^*(E)$ factor reported by Spillane et al. at low energies also exceeds the upper limits predicted by CC-M3Y+Rep calculations and constraints inferred from the $\rm ^{12}C+{}^{13}C$ reaction \cite{Spi2007}.
This discrepancy highlights the difficulty of reconciling experimental trends with existing theoretical models in this energy region, as most models do not include resonances, whereas in this case, theoretical models need to account for resonances to become reliable, as clearly pointed out by A. Bonasera and J.B. Natowitz \cite{Bon2020}.
The $\rm S^*(E)$ factor is calculated within the framework of a time-dependent Hartree-Fock-based classical model using the Feynman path-integral method \cite{Bon2020}, the result shows some agreement with the THM data \cite{Tum2018} when low-energy 0$^+$ resonances are added to the calculation, but in contrast to the modified $\rm S^*(E)$ factor with the Coulomb corrections \cite{Muk2019}. 
Furthermore, the theory proposed by Mukhamedzhanov predicts spectator particles distributed mainly at backward angles, which contradicts the forward-peaked distribution expected from the quasi-free mechanism.

These debates and doubts highlight the necessity for further experimental tests, whether direct or indirect. It is noteworthy that the Trojan-horse nucleus used in the Tumino experiment, $\rm ^{14}N= (^{12}C \oplus d)$, has a binding energy of $e_a$=10.27 MeV. Using a different Trojan-horse nucleus is therefore desirable to test the reliability of the method.

In this work, we apply the Trojan horse method based on DWBA, which incorporates Coulomb corrections. 
Considering that the binding energy of the Trojan Horse nucleus $\rm ^{14}N= (^{12}C \oplus \mathit{d})-10.27$ MeV is relatively high, and the binding energy of the spectator particle $d=(p+n)-2.225$ MeV is very small, there exists a competing cluster structure that is more prone to breakup: $\rm ^{14}N= (^{13}C \oplus \mathit{p})-7.55$ MeV. In this experiment, we chose $^{16}\mathrm{O}=( {}^{12}\mathrm{C} +\alpha)-7.16$ MeV as the Trojan Horse nucleus \cite{a16O1984, a16O1988} because of its lower binding energy and the fact that the spectator $\alpha$-particle is a typically tightly bound structure. Therefore, it can be theoretically expected to more readily produce a quasi-free breakup.

The application of $\alpha$ using as a spectator particle has been validated in previous THM experiments with the Trojan horse nucleus $\rm ^{6}Li= (\alpha \oplus \mathit{d})$ \cite{APJ1996,PRC2001,lcb2015,lcb2017} or $\rm ^{9}Be= (\alpha \oplus {}^{5}He)$ \cite{EPJA2020,EPJA2021}, with no evidence indicating that it causes significant distortion. 

By employing a copper beam-stopper foil, we successfully measured the spectator $\alpha$-particle around 0° for the first time in a THM experiment, a region densely populated by spectators under the quasi-free mechanism. The $\rm S^*(E)$ factor of $\rm {}^{12}C(^{12}C,\alpha)^{20}Ne$ in the Gamow energy region was extracted from the indirect measurement of the three-body reaction $\rm {}^{12}C(^{16}O,\alpha\alpha){}^{20}Ne$.

\section{\label{sec:method} Trojan Horse Method}

The Trojan horse method (THM) is a powerful indirect technique developed in experimental nuclear astrophysics \cite{Bau1986, Typ2000, Typ2003, Bau2004, Tri2014, Ber2018, epjrev2019}. It is used to extract the cross sections of charged-particle reactions at astrophysical energies, overcoming the suppression from the Coulomb barrier and electron screening effects that plague direct measurements. THM has been applied to study numerous astrophysically relevant reactions \cite{Tum2018, lcb2015, lcb2017, epjrev2019, prl2007, lac2007, prl2012, plb2015, apj2017, wen2008, wen2011, lcb2015a, wen2016, wang2024}.

The initial theoretical treatment of THM was based on the plane-wave impulse approximation (PWIA)  \cite{Bau1986}, which established the relationship between the two-body reaction cross section and an appropriate three-body one.  Later, S. Typel and G. Baur refined the method within the post-form distorted-wave Born approximation (DWBA) framework \cite{Typ2000, Typ2003, Bau2004}. This approach not only recovered a form similar to PWIA but also naturally incorporated the Coulomb barrier penetration factor, which compensates for the suppression of the two-body cross section \cite{Typ2003}.

\begin{figure}
\begin{center}
\includegraphics[width = 0.45\textwidth]{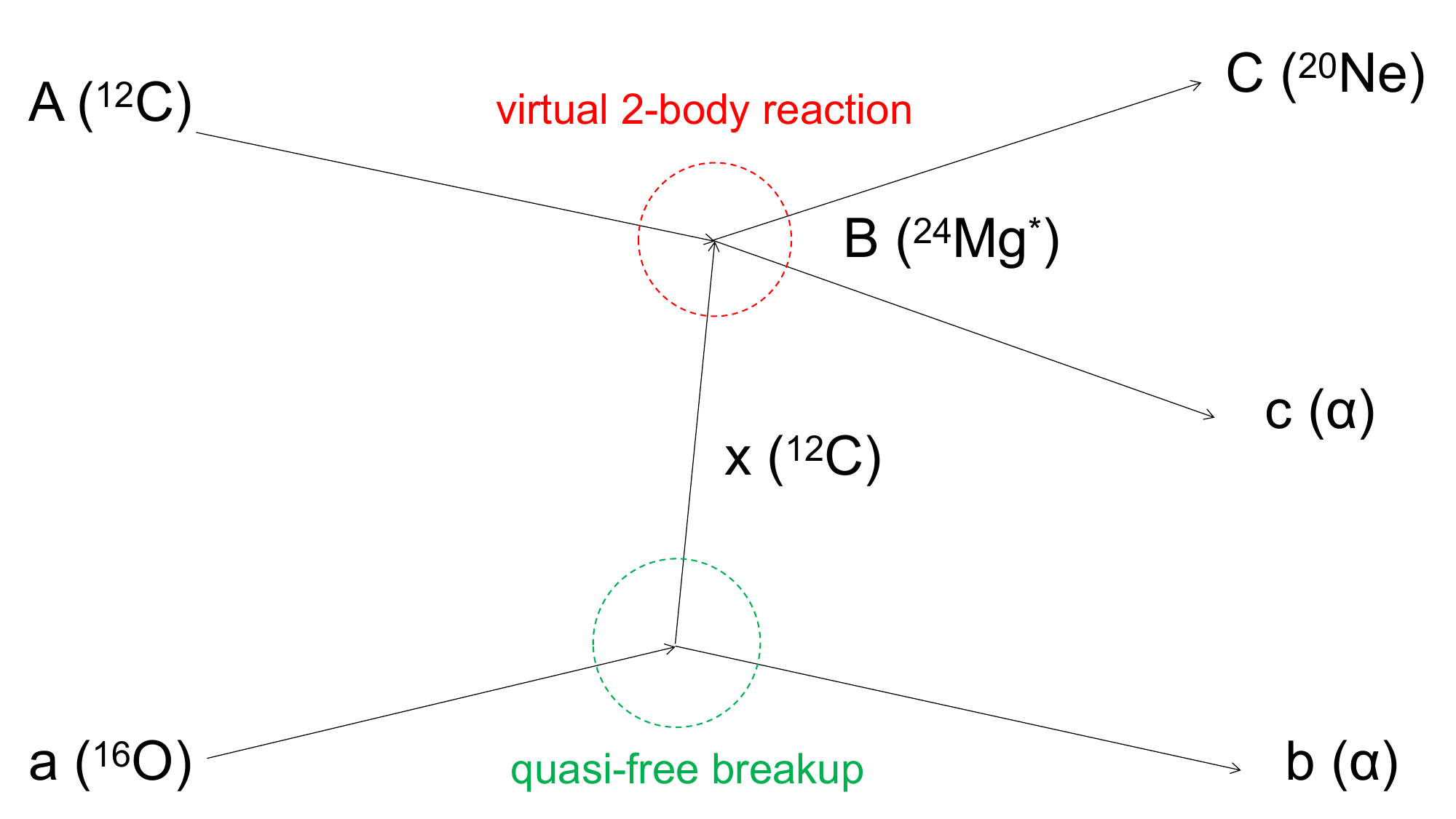}\\
\caption{(Color online) Schematic representation of the Trojan horse method. General: A(a,bc)C; this work:  $\rm {}^{12}C(^{16}O,\alpha \alpha){}^{20}Ne$ }
\label{figTHM} 
\end{center}
\end{figure}

As illustrated in Fig. {\ref{figTHM}}, the Trojan horse method relies on the quasi-free (QF) reaction mechanism. It allows the indirect determination of the low-energy cross section for a two-body reaction between charged particles Eq.(\ref{eq:two_body})
\begin{equation}\label{eq:two_body}
    A+x \rightarrow C+c
\end{equation}
from the measurement of a suitable three-body process Eq.(\ref{eq:three_body}) under quasi-free kinematic conditions at an energy above the Coulomb barrier:
\begin{equation}\label{eq:three_body}
    A+a \rightarrow C+c+b
\end{equation}
Here, the Trojan horse nucleus $a$ is considered to be predominantly composed of clusters $x$ and $b$ ($a=(x \oplus b)$).
Following the quasi-free breakup of nucleus $a$ induced by the interaction with nucleus $A$, the virtual two-body reaction occurs between the transferred particle $x$ and nucleus $A$,  while the other cluster $b$ acts as a spectator, retaining its original energy and momentum. 
Since the transferred nucleus $x$ is virtual, its energy and momentum do not satisfy the standard relation for a free particle, imparting a half-off-the-energy-shell (HOES) character to the A(x,c)C reaction.

The entrance channel energy $E_{Aa}$ is chosen to be above the Coulomb barrier to avoid suppression. 
However, the effective energy $E_{Ax}$ of the reaction between $A$ and $x$ can be very low, even sub-threshold. This is because part of $E_{Aa}$ is used to overcome the binding energy $\varepsilon_a=(m_x+m_b-m_a)c^2$ of $a=(x \oplus b)$, and the Fermi motion of $x$ inside $a$, characterized by $E_{xb}$, spans an energy region around the quasi-free point $E_{Ax}^{qf}$:
\begin{equation}\label{eq:EqEAx}
       E_{Ax}=E_{Ax}^{qf} \pm E_{xb}
\end{equation}
\begin{equation}\label{eq:EqEAxqf}
        E_{Ax}^{qf}=E_{Aa}\left(1-\frac{\mu_{Aa}}{\mu_{Bb}}\frac{\mu_{bx}^{2}}{m_x^2}\right)-\varepsilon_{a}
\end{equation}
where $E_{xb}$ is constrained by the wave number $\kappa_{xb} = (2\mu_{xb}\varepsilon_a)^{1/2}$ of the bound state. 

A single $E_{Aa}$ value corresponds to a range of $E_{Ax}$ values, allowing the extraction of two-body reaction data over a low-energy range from a three-body measurement at a fixed beam energy.  
By hiding the transferred particle  $x$ inside the Trojan-horse nucleus $a$, it can be brought into the nuclear interaction region without Coulomb suppression or electron screening effects, enabling the measurement of the bare nuclear astrophysical S(E) factor.
The energy of the two-body reaction is derived from the measured outgoing particle energy $E_{Cc}$ : 
\begin{equation}\label{eq:EqEAxCc}
       E_{Ax}=E_{Cc}-Q_2
\end{equation}

Finally, after some theoretical simplification (such as surface approximation)\cite{Typ2003}, the three-body reaction cross section can be factorized:

\begin{equation}\label{eq:sec3}
\frac{d^3\sigma}{dE_{Cc}d\Omega_{Bb}d\Omega_{Cc}} = K_F \cdot |W|^2  \cdot  \frac{d\sigma}{d\Omega}^{TH} 
\end{equation}

$K_F$ is a kinematic factor containing the final-state phase-space factor, dependent on masses, momenta, and angles.
\begin{equation}\label{eq:kf}
 K_F = \frac{\mu_{Aa}\mu_{Bb}\mu_{Cc}}{(2\pi)^5 \hbar^6} \frac{k_{Bb}k_{Cc}}{k_{Aa}} \frac{16\pi^2}{k_{Ax}Q_{Aa}} \frac{v_{Cc}}{v_{Ax}}
\end{equation}

$|W|^2$ is the momentum distribution of the spectator $b$ inside the Trojan-horse nucleus $a$,  
its amplitude is:

\begin{equation}\label{eq:W}
 W(Q_{Bb}) = - \left(\varepsilon_a + \frac{\hbar^2 Q_{Bb}^2}{2\mu_{xb}} \right) \left\langle \exp(iQ_{Bb} \cdot r_{xb})\phi_x \phi_b | \phi_a  \right\rangle 
\end{equation}

${\sigma}^{TH}$is the HOES cross section for $A+x\rightarrow C+c$,  which can be expressed as:
\begin{equation}\label{eq:sec2}
\frac{d\sigma_l}{d\Omega}^{TH} = P_l  \cdot \frac{d\sigma_l}{d\Omega}({Ax\rightarrow Cc)}
\end{equation}
where $\frac{d\sigma_l}{d\Omega}({Ax\rightarrow Cc)}$  is the on-energy-shell cross section for partial wave $l$, and $P_l$ is defined as:
$$P_l = k^2 R^2 [F_l^2 + G_l^2] z_l^2$$
where $F_l$ and $G_l$ are the Coulomb wave functions, and $z_l$ is the Riccati–Bessel function.
It is worth noting that the $P_l$ defined here differs from the standard definition of the Coulomb penetration factor ($P_l = k R / [F_l^2 + G_l^2]$). Instead, it serves as a compensation factor for the Coulomb penetration effect. Its purpose is to counteract the sharp decrease in the cross-section at low energies — caused by the Coulomb barrier in on-energy-shell two-body cross section — thereby rendering the $\sigma^{TH}$ variation with energy more gradual.

Thus, the three-body reaction cross section for partial wave $l$ can be expressed as:
\begin{equation}\label{eq:sec3all}
\frac{d^3\sigma}{dE_{Cc}d\Omega_{Bb}d\Omega_{Cc}} = K_F \cdot |W|^2  \cdot  P_l  \cdot \frac{d\sigma_l}{d\Omega}({Ax\rightarrow Cc)} 
\end{equation}

The energy relationship Eq.(\ref{eq:EqEAxqf}) guides the experimental beam energy selection (typically setting $E_{Ax}^{qf}$ near $E_G$), and the cross section relationship Eq.(\ref{eq:sec3all}) is used to extract the relevant two-body reaction cross section from the measured three-body data  after selecting quasi-free events.
The $\rm S^*(E)$ factor is then determined from its definition in Eq.(\ref {eq:SEm}).

\section{\label{sec:exp} Experimental Setup}

\begin{figure*}
\begin{center}
\includegraphics[width = 0.95\textwidth]{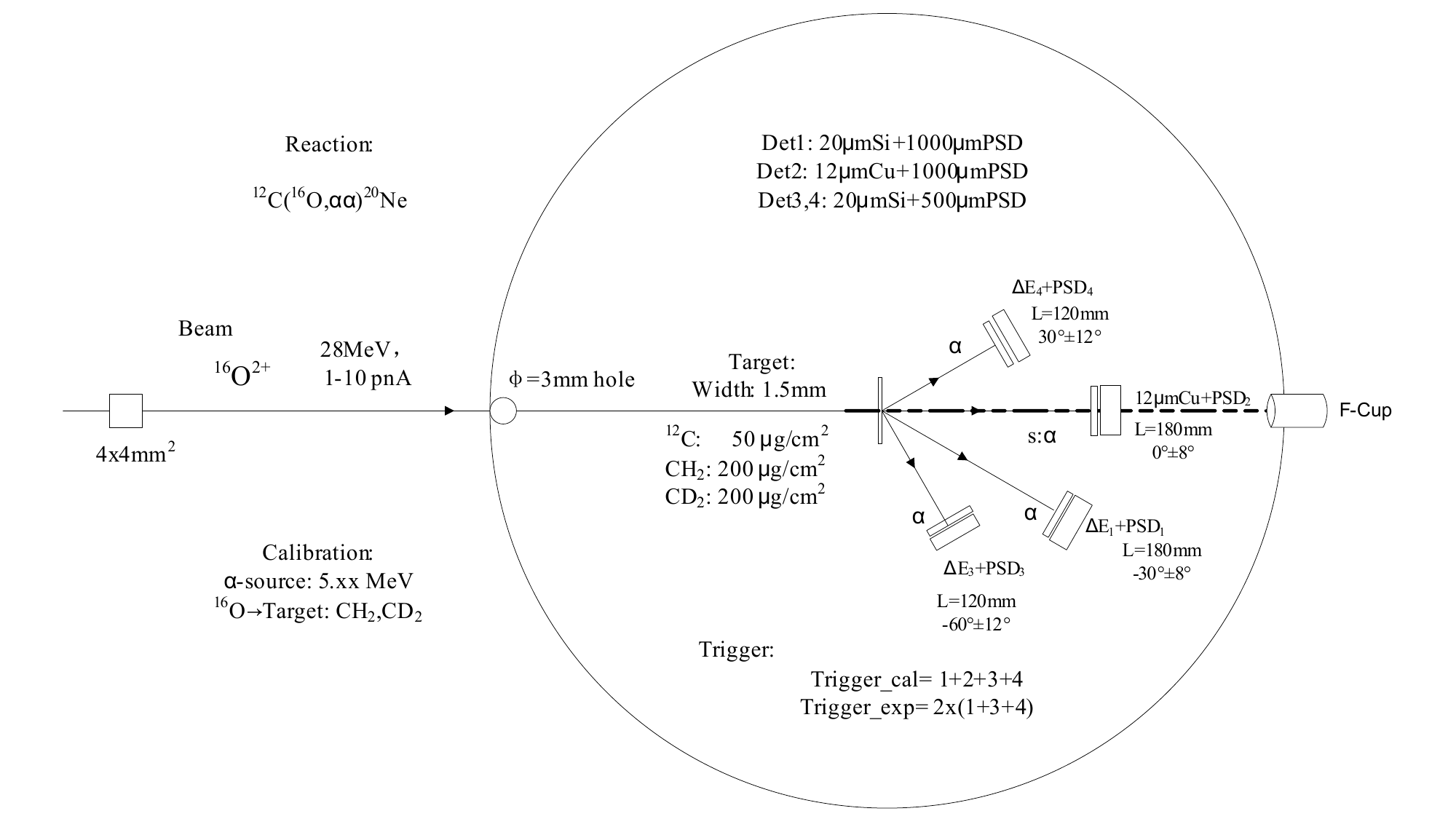}\\
\caption{Schematic diagram of the experimental setup}
\label{figSetup} 
\end{center}
\end{figure*}



The experiment was performed at the HI-13 Tandem Accelerator  Laboratory of China Institute of Atomic Energy (CIAE). A schematic of the experimental setup is shown in Fig. {\ref{figSetup}}. A carbon target was bombarded with a 28 MeV $\rm ^{16}O^{2+}$ beam with a current of approximately 10 particle-nanoamperes (pnA) to induce the three-body nuclear reaction $\rm ^{12}C(^{16}O, \alpha \alpha)^{20}Ne$. The carbon target had a thickness of 50 $\rm \mu gcm^{-2}$ and a width of 1.5 mm ( to minimize the angular uncertainty caused by a large or misaligned beam spot, we employed a slender, vertical stripe target with a height of 1 cm and a width of 1.5 mm ). The target frame  also held 200 $\rm \mu gcm^{-2}$  $\rm CH_2$ and $\rm CD_2$ targets for calibration purposes. Online detector calibration was performed using the $\rm ^{1}H(^{16}O,p)^{16}O$, $\rm ^{2}H(^{16}O,d)^{16}O$ and $\rm ^{2}H(^{16}O,\alpha)^{14}N$ reactions, supplemented by offline calibration with an $\alpha$-source.

As shown in Fig. {\ref{figSetup}}, three $\rm \Delta E-E_r$ telescope detectors were placed on each side of the beam line at 30° and 60° to measure and identify light particles from the reaction. 
The $\rm \Delta E$ detector was a 20 $\rm \mu m$ thick single side silicon micro-strip detector (SSSD). The $\rm E_r$ detector was a one-dimensional position-sensitive detector (PSD) with thickness of 500 or 1000 $\rm \mu m$ and an active area of $\rm 50 \times 10 ~ mm^2$.  The PSD achieved a position resolution of about 0.3 mm, and an energy resolution of better than 45 keV for 5.5 MeV $\alpha$ source. A cooling system was used to effectively reduce the thermal noise of the detectors and integrated preamplifiers.

A single 1000 $\rm \mu m$ thick PSD, covered by a 12 $\rm \mu m$ Cu beam-stopper foil, was positioned around 0° (0° ± 8°) relative to the beam line to measure the spectator $\alpha$-particles from the quasi-free reaction. 
The trigger of the DAQ system is set as 'Trigger\_cal = det1 + det2 + det3 + det4' for the detector calibration runs, and 'Trigger\_exp = det2 x (det1 + det3 + det4)' for the measurement runs.
The energy and angle information of particles were obtained through coincidence measurements between the telescope detectors and the 0° PSD.

The use of the beam-stopper foil, for the first time in THM experiments, was crucial for measuring spectator particles around 0°, where they are most concentrated under the quasi-free mechanism. This approach offers several advantages:

(1), It protects the downstream detector from potential damage caused by heavy ions such as C, N and O generated by strong beam scattering and reactions.

(2), It blocks heavy ions while allowing light particles ($p$, $d$, $\alpha$) to pass through, enabling detection at very forward angles around 0°.

(3), By eliminating background from beam scattering, it allows for higher beam intensities, improving beam-time utilization and data acquisition rates.

(4), By targeting the angular region around 0° with the highest density of spectator particles, it maximizes the yield of quasi-free events for a given beam time, enhancing statistical precision and reducing errors.

Neglecting non-uniformities in the beam-stopper foil thickness, a Geant4 simulation indicated that the energy resolution degradation caused by the 12 $\rm \mu m$ Cu foil was equivalent to a detector resolution of $\sigma$ = 21.25 keV (FWHM = 50 keV).

\section{\label{sec:data} Data Analysis and Result Discussion}

\subsection{Selection of three-body reaction events}

\begin{figure}
\begin{center}
\includegraphics[width = 0.45\textwidth]{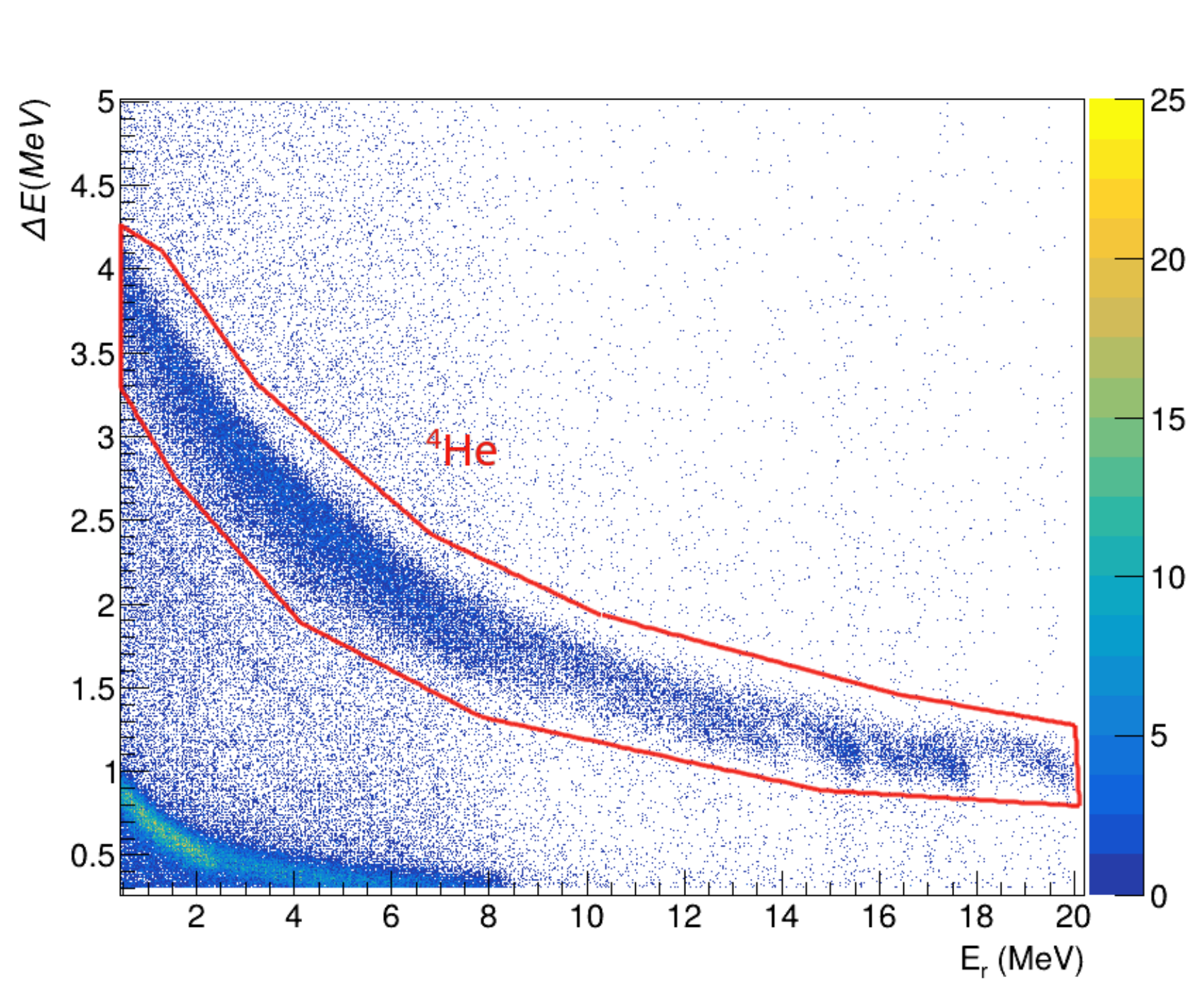}\\
\caption{(Color online) Particle identification using $\rm \Delta E-E_r$ spectrum to select $p$, $d$, or $\alpha$. }
\label{figdEE} 
\end{center}
\end{figure}

Following detector calibration, light particles ($p$, $d$, $\alpha$) were identified using the $\rm \Delta E-E_r$ spectrum, as shown in Fig. {\ref{figdEE}}.

Alpha particles originating from the $\rm ^{12}C(^{16}O, \alpha \alpha)^{20}Ne$ three-body reaction were selected from the $\rm \Delta E-E_r$ telescopes in coincidence with a spectator $\alpha$-particle  detected by the 0° PSD. 
Assuming the undetected third particle is $\rm ^{20}Ne$, the event kinematics were reconstructed, allowing the calculation of the third particle's energy, angle, and momentum. The experimental Q value of the three-body reaction,  $\rm Q_3=E_1+E_2+E_3-E_0$, as well as the relative energy $\rm E_{12}$, $\rm E_{23}$, $\rm E_{13}$, were then determined.

\begin{figure}
\begin{center}
\includegraphics[width = 0.45\textwidth]{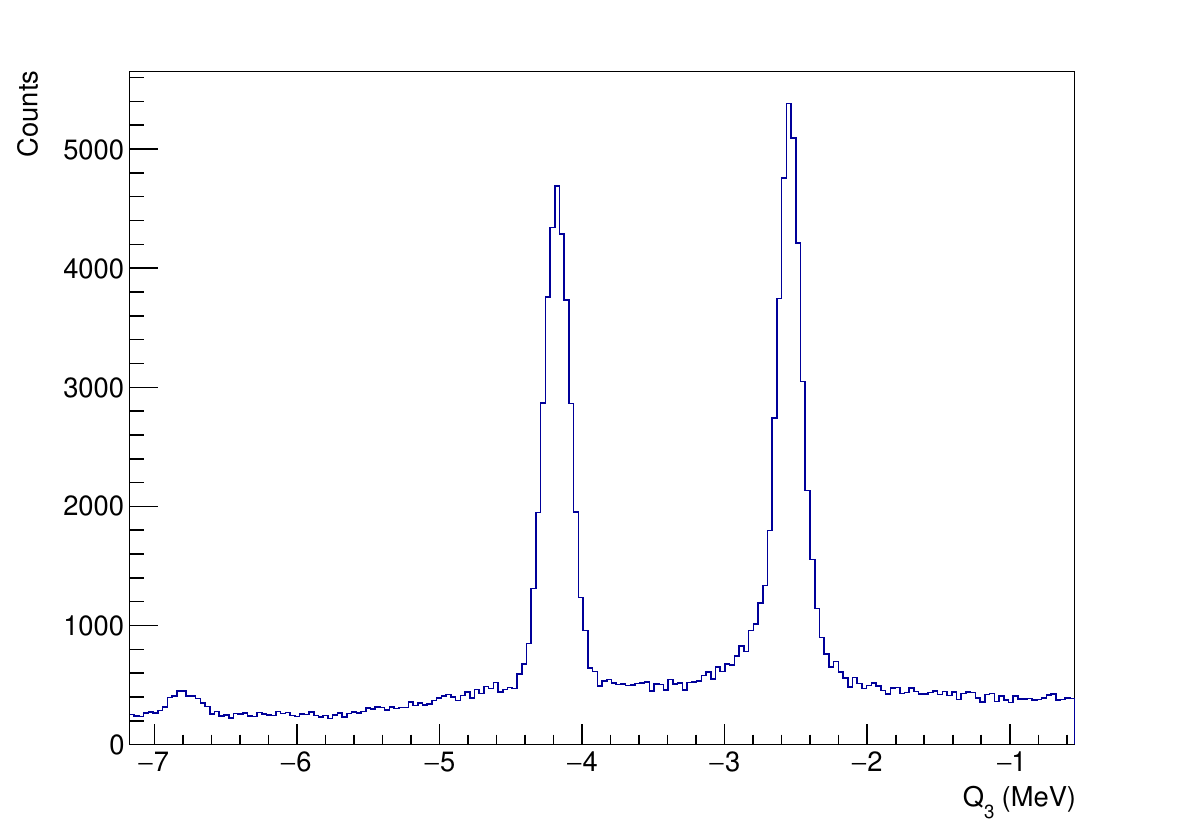}\\
\caption{(Color online) Spectrum of the experimental $Q_3$ values used to select events of the three-body reaction $\rm ^{12}C(^{16}O, \alpha \alpha)^{20}Ne$: $Q_3$ = -2.545 MeV for $\alpha_0$ channel, and $Q_3$ = -4.175 MeV for $\alpha_1$ channel. }
\label{figQ3} 
\end{center}
\end{figure}

As shown in Fig. {\ref{figQ3}}, events corresponding to the three-body reaction $\rm ^{12}C(^{16}O, \alpha \alpha)^{20}Ne$ were selected from the experimental $\rm Q_3$ spectrum by applying cuts around the theoretically expected Q-values: $\rm Q_3$= -2.545 MeV for the $\alpha_0$ channel and $\rm Q_3$ = -4.175 MeV for the $\alpha_1$ channel.

\subsection{Quasi-free condition test}

After selecting three-body reaction events for the $\alpha_0$ (or $\alpha_1$) channel via the $\rm Q_3$ cut, it is essential to verify that these events satisfy the quasi-free condition.

The momentum distribution of the spectator $\alpha$-particle is highly sensitive to the reaction mechanism. It retains the shape of its momentum distribution inside $\rm ^{16}O$ only if the breakup is quasi-free. Therefore, agreement between the experimental and theoretical momentum distribution shapes is a strong signature of the quasi-free mechanism.

In practice, in THM experiments, by rationally designing the detector configuration, we can preferentially detect regions where quasi-free events are dominant (around 0 degree in this work). Furthermore, kinematic selection conditions (for instance, by utilizing THM simulations to select the energy-angle correlation regions for each outgoing particle, the angular correlation and energy correlation regions between two outgoing particles) are applied during data analysis to reject interfering events. However, because quasi-free events and other interfering events often partially overlap in kinematics, complete separation cannot be achieved solely through kinematic conditions. Therefore, comparing the experimental data with the theoretical curve—and selecting the momentum region where they are in good agreement as the cut window for subsequent analysis—is one of the core methods to ensure data quality.

The momentum distribution of the $\alpha$-particle inside the $\rm ^{16}O$ ground state is essentially the Fourier transform of the coordinate-space wave function — which is obtained by solving the Schrödinger equation with a Woods-Saxon potential — into momentum space. 
This distribution exhibits a peak strictly at p=0 and follows a Gaussian-like profile for a Tojan horse nucleus with $l=0$.

\begin{figure}
\begin{center}
\includegraphics[width = 0.45\textwidth]{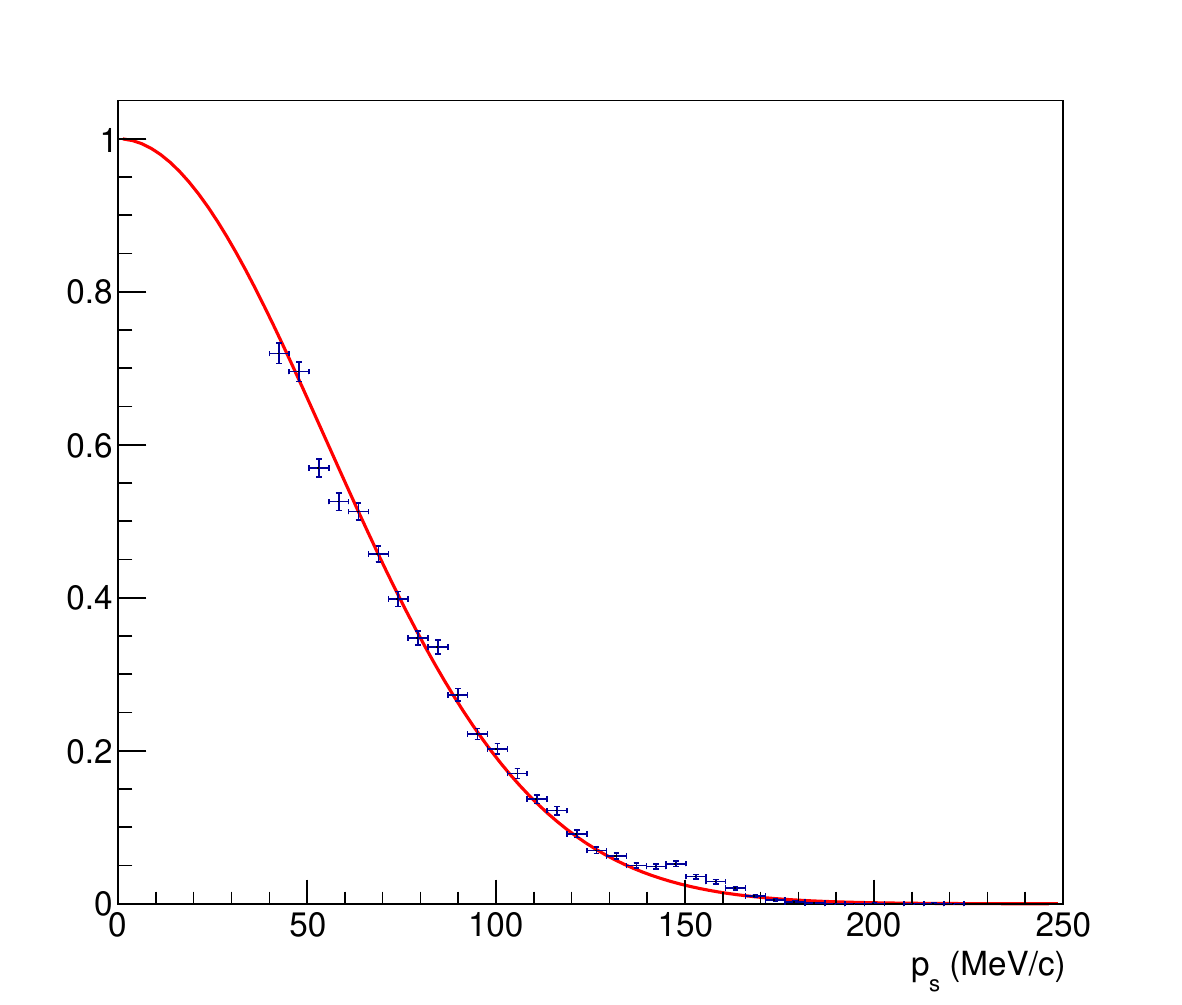}\\
\caption{Comparison of the experimentally measured momentum distribution of the spectator $\alpha$-particle (blue points) with the theoretical momentum distribution for an $\alpha$-cluster inside $\rm ^{16}O$ (red solid line).}
\label{figps} 
\end{center}
\end{figure}

Fig. {\ref{figps}} compares the experimental center-of-mass momentum distribution of the spectator $\alpha$ (for the $\alpha_0$ channel) with the theoretical momentum distribution for an $\alpha$-cluster inside $\rm ^{16}O$. 

The theoretical curve (red solid line) was obtained from the depth of a Woods-Saxon potential, adjusted to reproduce the experimental ground-state binding energy (7.16 MeV) for the $\rm {}^{16}O=({}^{12}C \oplus \alpha)$ configuration, while its geometrical parameters (radius, diffuseness) were fixed to standard values from the literature. The parameters we used here as: $R$ = 4.65 fm, $a$ = 0.65 fm and $V_0$ = 32.45 MeV.

The experimental data points are normalized to the theoretical curve. The good agreement in the range of 50-130 MeV/$c$ confirms the dominance of the quasi-free mechanism, the proportion of interfering events is negligible. A cut of $50<p_s<130$ MeV/$c$ was applied to select quasi-free events for subsequent analysis.
The absence of data points $p_s$ < 50 MeV/$c$ is due to the cutoff of the detection threshold for low-energy $\alpha$-particles, which must pass through the beam-stopper foil and the dead layer of the PSD, and the noise-rejection cuts applied during data analysis.

Furthermore, since the Trojan-horse nucleus $\rm {}^{16}O=({}^{12}C \oplus \alpha)$ is used as the beam, the spectator $\alpha$-particles are expected to be concentrated at small angles around 0° along the beam direction in a quasi-free reaction. Thus, the measured angular distribution of the spectator $\alpha$ around 0° provides another critical test of the quasi-free condition.

\begin{figure}
\begin{center}
\includegraphics[width = 0.45\textwidth]{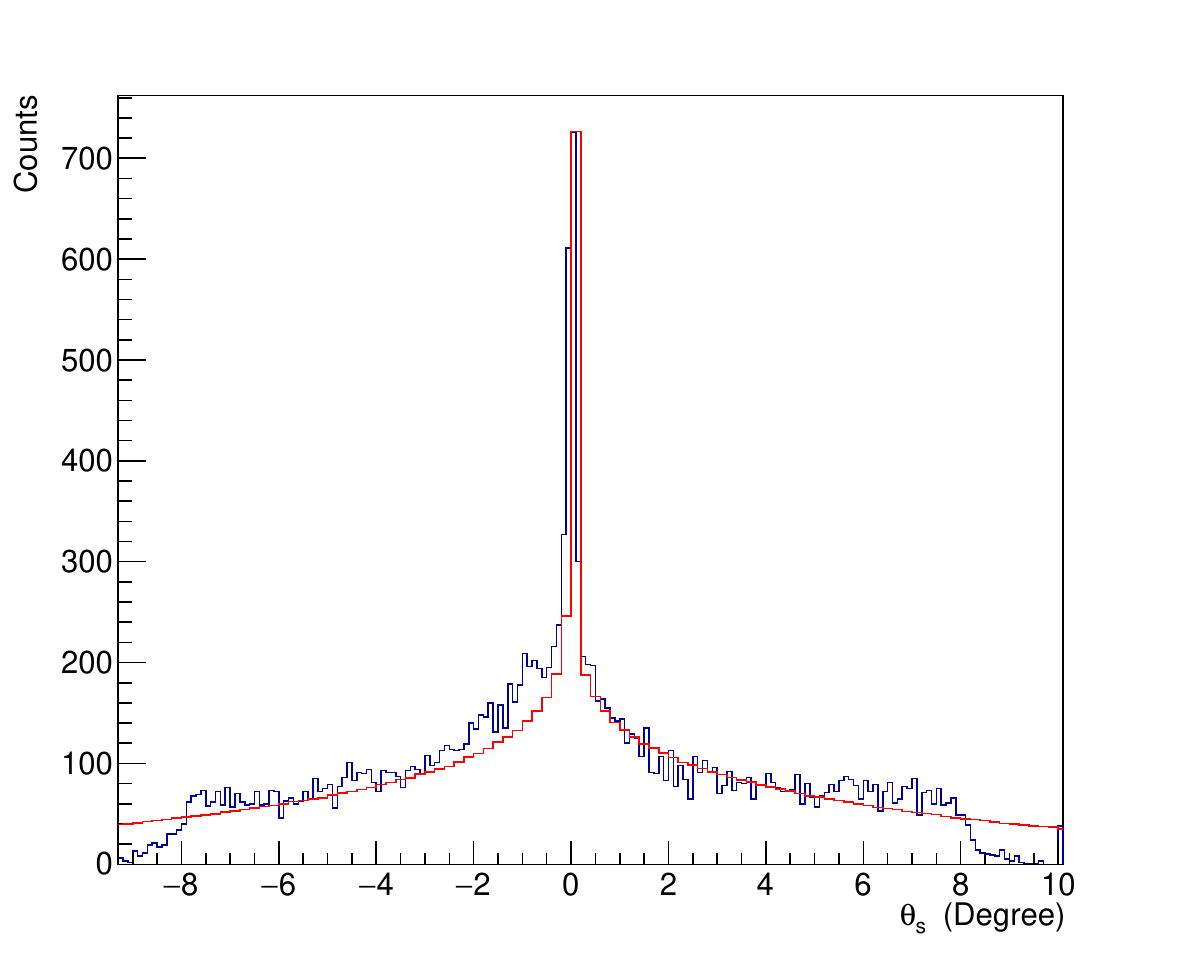}\\
\caption{(Color online) Angular distribution of the spectator $\alpha$-particle around 0°: comparison between the experimental measurement (blue line) and the simulation of the quasi-free process (red line).}
\label{figthps} 
\end{center}
\end{figure}

As shown in Fig. {\ref{figthps}}, the angular distribution of the spectator $\alpha$ around 0° (blue line) was extracted by applying the cuts of $\Delta E-E_r$, $Q_3$, and $p_s$. Its comparison with a simulation of the THM quasi-free process (red line) demonstrates that the selected events are predominantly from the quasi-free process.

\subsection{S-factor of the two-body reaction}

In nuclear astrophysics, the astrophysical $S(E)$ factor is introduced to eliminate the main energy dependence of the charged particle reaction cross section at astrophysical energies, which drops sharply with energy reduction due to Coulomb barrier penetration. 

The  $S(E)$ factor is defined as:
\begin{equation}\label{eq:SE}
    S(E)= \sigma (E) E \exp(2\pi\eta)
\end{equation}
where $E$ is the incident energy in the center-of-mass system, $\sigma(E)$ is the energy-dependent cross section and $\rm exp(2\pi \eta)$ is the inverse of the Gamow factor, with $\eta$ the Somerfield parameter, $\eta=Z_1 Z_2 e^2 /4 \pi \epsilon_0 \hbar v$. 

For $s$-wave non-resonant reactions, the $S(E)$ factor is nearly independent of energy and it is the conventional quantity used to extrapolate to low energies.

The modified astrophysical S-factor, $S^*(E)$, is commonly used in carbon burning studies \cite{Pat1969}:

\begin{equation}\label{eq:SEm}
    S^*(E)= \sigma (E) E \exp(87.21E^{-1/2} + 0.46 E)
\end{equation}
Here, the exponential term includes a correction from the second term in the Coulomb barrier approximation, with the numerical factor 0.46 corresponding to the size parameter $g = 1/3(\mu R_0^3/2Z_1Z_2)^{1/2}$.

After selecting the quasi-free three-body events as described, the cross section relationship of Eq.(\ref{eq:sec3all}), derived from the post-form DWBA theory of THM \cite{Typ2003}, was used to extract the cross section for $\rm {}^{12}C({}^{12}C,\alpha){}^{20}Ne$ from the measured three-body $\rm ^{12}C(^{16}O, \alpha \alpha)^{20}Ne$ cross section. The $S^*(E)$  factor was then calculated using its definition in Eq.(\ref{eq:SEm}).

The Coulomb penetration compensation factor $P_l$ in our analysis was calculated primarily for the $l=0$ partial wave. We evaluated the contributions from $l=1$ and $l=2$ waves and found their influence on the overall shape of the $S^*(E)$ factor to be negligible compared to the dominant $l=0$ component. 

\begin{figure}
\begin{center}
\includegraphics[width = 0.475\textwidth]{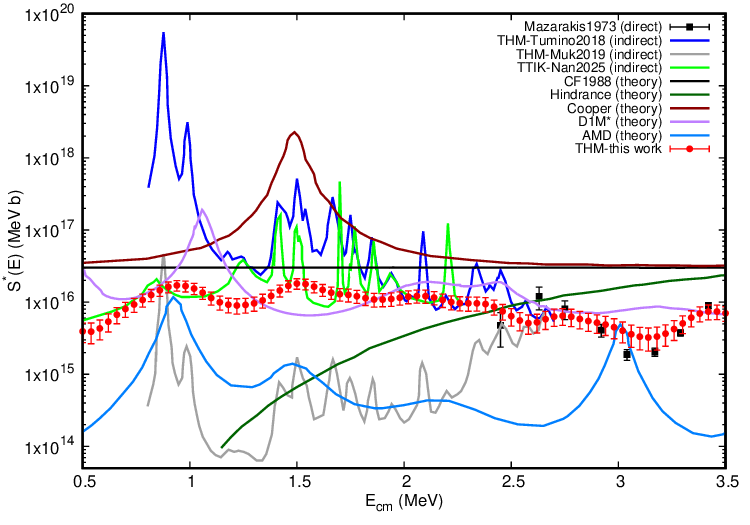}\\
\caption{(Color online) Comparison of the $\rm S^*(E)$ factor of $\rm {}^{12}C({}^{12}C,\alpha_0){}^{20}Ne$ extracted by THM in this work with other experimental measurements and theoretical curves. Note: the theoretical curves (CF1988, Cooper, Hindrance) represent the total $\rm S^*(E)$ factor. }
\label{figSa0} 
\end{center}
\end{figure}

\begin{figure}
\begin{center}
\includegraphics[width = 0.475\textwidth]{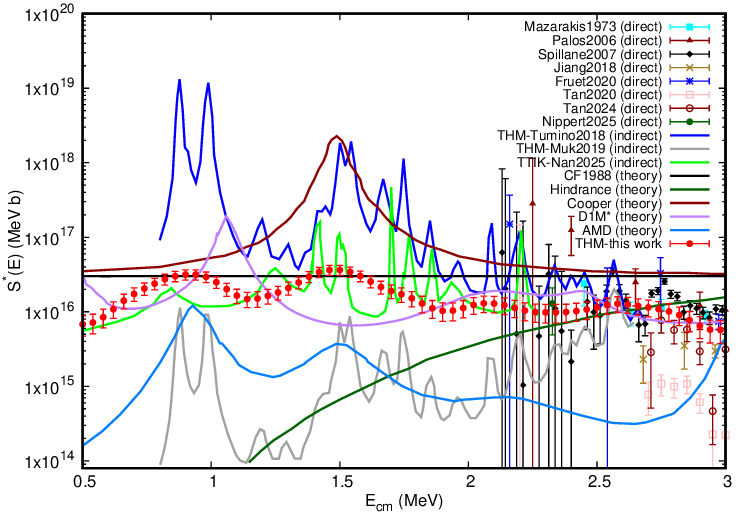}\\
\caption{(Color online) Comparison of the $\rm S^*(E)$ factor of $\rm {}^{12}C({}^{12}C,\alpha_1){}^{20}Ne$ extracted by THM in this work with other experimental measurements and theoretical curves.  Note: the theoretical curves (CF1988, Cooper, Hindrance) represent the total $\rm S^*(E)$ factor. }
\label{figSa1} 
\end{center}
\end{figure}

The absolute scale of the THM  $S^*(E)$ factor must be normalized to direct measurement data. 
For the $\alpha_0$ channel, normalization was performed in the 2.5-3.5 MeV energy range using the direct data from Mazarakis 1973 \cite{Maz1973}. 
For the $\alpha_1$ channel, normalization was carried out in the 2.5-3.0 MeV range using direct data sets of \cite{Spi2007,Maz1973,Bar2006,Jcl2018,Fru2020,Tan2020,tan2024,nip2025}. 
The resulting $S^*(E)$ spectra are shown in Fig. \ref{figSa0} and \ref{figSa1}.

\begin{figure}
\begin{center}
\includegraphics[width = 0.475\textwidth]{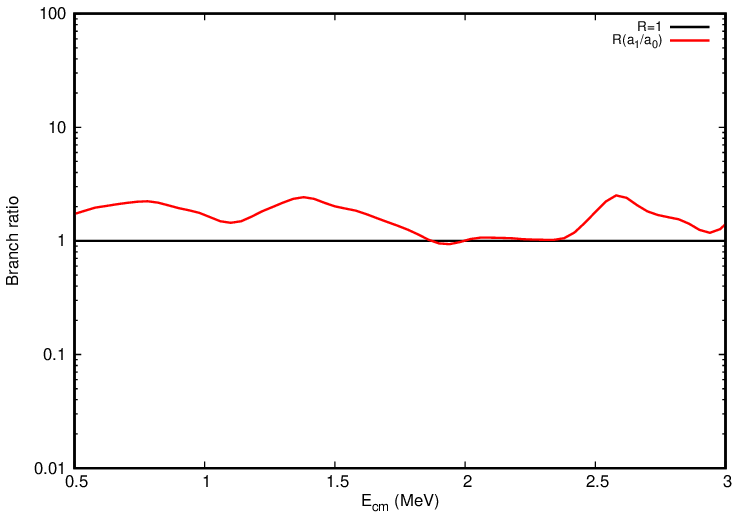}\\
\caption{(Color online) Branching Ratio $R(\alpha_1/\alpha_0)$ extracted from the THM data in this work.}
\label{figRa} 
\end{center}
\end{figure}

The branching ratio $R(\alpha_1/\alpha_0)$ extracted from our THM data is presented in Fig. \ref{figRa}. 
The branching ratios exhibit a fluctuating variation, and their absolute values are affected by the uncertainty of the normalization coefficients, particularly for the $\alpha_1$ channel, as the direct measurements differ significantly from one another. Considering the effect of the experimental energy resolution on peak broadening, the actual fluctuation in the branching ratios may be even more pronounced.

It should be noted that we opted not to use an R-matrix analysis in this work for two main reasons. First, there are significant discrepancies in the reported resonance parameters and the number of included levels between different R-matrix studies of $\rm ^{12}C+{}^{12}C$ reaction \cite{Tum2018,wyb2025}, particularly for states with disputed spin-parity. Second, and most critically, the broadened lineshapes in our data, primarily a consequence of our experimental energy resolution, mean that forcing an R-matrix fit to the data would not yield reliable or unique resonance parameters.
Although the resonance peaks are significantly broadened and the resonance strength is weakened due to the combined effects of experimental energy resolution and the limitations of the surface approximation within the DWBA framework, our approach provides a model-independent (for levels selection with parameters) determination of the gross structure of the $S^*(E)$ factor. This is particularly valuable for constraining the overall trend at low energies (flat, rising, or falling), which is a matter of ongoing debate. The data honestly reflect the resolving power achievable with our detector setup.

The vertical ($S^*(E)$) uncertainty is dominated by statistical errors, while the horizontal ($ E_{cm}$) uncertainty is primarily governed by the overall energy resolution of the detection system.
The energy resolution of the experimental setup is the key factor limiting the precision of the extracted resonance parameters. The overall resolution in the center-of-mass system is estimated to be $\sigma \sim$ 110 keV. This broadening stems from several contributors: the inherent energy resolution of the PSD and SSSD detectors, uncertainties in the energy calibration and kinematic reconstruction process, and energy straggling in the beam-stopper foil. A comprehensive evaluation suggests that non-uniform thickness of the beam-stopper foil is likely the dominant source degrading the final energy resolution.

\subsection{Result discussion}

The $\rm S^*(E)$ factor of $\rm {}^{12}C(^{12}C,\alpha)^{20}Ne$ reaction in the Gamow energy region has been successfully extracted via the DWBA-based THM applied to the $\rm {}^{12}C(^{16}O,\alpha\alpha){}^{20}Ne$ three-body reaction. 

The results are compared with direct experimental data \cite{Spi2007,Maz1973,Bar2006,Jcl2018,Fru2020,Tan2020,tan2024,nip2025}, previous indirect THM results \cite{Tum2018}, the Coulomb-corrected THM data \cite{Muk2019}, recent indirect measurements using the thick-target inverse kinematics (TTIK) method \cite{wyb2025}, and various theoretical predictions \cite{CF1988, Hin2007, Cop2009, AMD2021, DIM2024}, as shown in Fig. \ref{figSa0} and \ref{figSa1}. 
The three theoretical model curves (CF1988, Cooper, Hindrance) are calculated for the total $\rm S^*(E)$ factor. Consequently, they are significantly higher than the experimental data for individual branching channels and can only serve as a qualitative guide for the overall trend in this context.
Key observations from this comparative analysis are as follows:

(1) \textbf{General Behavior of the $S^*(E)$ Factor:} 

\textit{General Behavior:} Overall, the $S^*(E)$ factor is characterized by a series of resonant structures superimposed on a relatively smooth, non-resonant platform. The superposition of these low-energy resonances leads to a gradual rising trend as the energy decreases.
However, influenced by the experimental energy resolution and the DWBA theoretical treatment employed, the resonances measured in this work are significantly broader than those reported in \cite{Tum2018,wyb2025}, where R-matrix fitting was employed to deconvolve the underlying states.
Consequently, what might be an original narrow resonance appears broadened, or alternatively, a cluster of closely spaced narrow resonances is merged into a single, wider peak, resulting in weakened resonance strengths.
Therefore, our extracted $S^*(E)$ factor can be regarded as the lower limit of the true $S^*(E)$ factor, demonstrating the overall trend of $S^*(E)$ changes with energy.

\textit{Resonance Structures:} 
Our results confirm the presence of resonances in both the $\alpha_0$ and $\alpha_1$ channels within the energy range from 0.5 to 3.0 MeV (peaks around 0.8, 1.5, 2.1 and 2.7 MeV), particularly inside the Gamow window around 1.5 MeV, as predicted by the Cooper \cite{Cop2009} and previously observed by Tumino2018 \cite{Tum2018}. 

(2) \textbf{Comparison with other Experimental Data and Theoretical Models:} 

\textit{Comparison with Indirect Data of THM and TTIK:}
Our results generally support the conclusions of the earlier indirect THM work \cite{Tum2018} and the TTIK2025 study \cite{wyb2025}: resonances do exist in the astrophysical energy region, particularly around 1.5 MeV.  
Without considering the detailed resonance structures, the overall trend of our results is qualitatively consistent with both the THM data of Tumino2018 \cite{Tum2018} and the TTIK2025 data \cite{wyb2025}.  
Although our data appear closer to the TTIK2025 dataset \cite{wyb2025} in the low-energy region (0.5-1.1 MeV), this should be taken only as an indication of the general trend and cannot be used to make a definitive distinction between the different datasets, due to the limited experimental energy resolution and the inherent limitations of the DWBA framework.

\textit{Comparison with the Modified THM-Muk2019:}
The Coulomb-corrected data THM-Muk2019 \cite{Muk2019} are significantly lower than  THM2018\cite{Tum2018}, TTIK2025\cite{wyb2025}, and our new measurements. Moreover, they generally exhibit a trend where resonance structures are superimposed on a non-resonant background that decreases with decreasing energy. In particular, within the energy range of 1.2–2.5 MeV, the variation trends show significant differences.

\textit{Comparison with Theoretical Models:}
In comparison with theoretical models, our data exhibit a relatively flat trend, which is similar to CF1988\cite{CF1988}. However, a resonant peak near 1.5 MeV is observed, consistent with the prediction of Cooper\cite{Cop2009}. The overall trend is entirely distinct from that of the Hindrance model\cite{Hin2007}. Quantitatively, our data are closest in magnitude to the D1M* curve\cite{DIM2024}, but the positions of the resonant peaks do not align well. In contrast, a qualitative comparison based on the positions of the resonant structures shows better agreement with the AMD model\cite{AMD2021}. However, because the AMD calculations do not include non-resonant contributions and possible interference effects, their overall magnitude is lower.

(3) \textbf{Evaluation of Low-Energy Hindrance:} 

Regarding the assessment of low-energy hindrance theoretical models \cite{Hin2007}, all existing experimental measurements above 1 MeV largely rule out the possibility of a strong hindrance effect. Although the $S^*(E)$ factor shows a slow decreasing trend below 0.8 MeV, this is primarily caused by the downward tail of the resonance peak near 0.8 MeV. 
Furthermore, the hindrance model predicts that the hindrance effect sets in at around 4 MeV and causes the $S^*(E)$ factor to decrease dramatically as the energy decreases.
Therefore, it is not possible to conclude that a hindrance effect exists at low energies. 


\section{\label{sec:sum} Summary}

In summary, resonances in the $\rm ^{12}C+^{12}C$ reaction within the Gamow window are critical for modeling various astrophysical scenarios but remain inaccessible to precise direct measurement.

While the indirect measurements of THM \cite{Tum2018} and TTIK \cite{wyb2025} have successfully explored these resonances, the number of reported resonances appears more than expectation from the identical $\rm ^{12}C+^{12}C$ two-boson system \cite{wyb2025}. 

In this work, we selected $\rm {}^{16}O=({}^{12}C \oplus \alpha)$ as the Trojan-horse nucleus due to its lower binding energy, which favors quasi-free reactions. The indirect measurement of $\rm {}^{12}C(^{16}O,\alpha\alpha){}^{20}Ne$ was performed at the HI-13 Tandem Accelerator in CIAE.

A key innovation was the use of a copper beam-stopper foil, enabling, for the first time in a THM experiment, the detection of spectator $\alpha$-particles within the small angular range around 0° where they are most abundant.

The $\rm S^*(E)$ factor of $\rm {}^{12}C(^{12}C,\alpha)^{20}Ne$ in the astrophysical energy region was extracted from the three-body reaction using the DWBA-based THM, without employing R-matrix fitting. 

Overall, the $S^*(E)$ factor is characterized by a series of resonant structures superimposed on a relatively smooth, non-resonant platform. 

Our results confirm the existence of resonances within the Gamow window around 1.5 MeV found by Tumino2018 \cite{Tum2018}. However, the resonances are broadened due to the experimental energy resolution and the limitations of the surface approximation within the DWBA framework.

Without considering the details of the resonance structures, the overall trend of our results is qualitatively in reasonable agreement with the THM-Tumino2018\cite{Tum2018} and TTIK2025\cite{wyb2025} data, but differs significantly from the trend of the Modified-THM-Muk2019\cite{Muk2019} data.

We observe no evidence for hindrance effect \cite{Hin2007} in the low energy range above 1 MeV. 
Although the $S^*(E)$ factor shows a slow decreasing trend below 0.8 MeV, this is primarily caused by the downward tail of the resonance peak near 0.8 MeV. 

The energy resolution of the experimental setup and the DWBA theoretical approximations is the primary factor limiting the precision of these results and must be carefully considered in future research.

\appendix
\section{Acknowledgments}

We thank the Nuclear Reaction Group in CIAE for their kind help during the experimental preparation and measurement.  
We gratefully acknowledge the staff of the HI-13 Tandem Accelerator Laboratory for providing the beam and targets.
We also thank Prof. Weiping Liu, Prof. Xiaodong Tang, Prof. Zhihong Li, Prof. Bing Guo, Prof. Youbao Wang, and Prof. Shuhua Zhou for fruitful discussions. 

This work is supported by the National Natural Science Foundation of China (Grants No. 12075031, 12275360) and the Natural Science Foundation of Beijing Municipality (Grant No. 1222022).


\section{Declaration of competing interest}

The authors declare that they have no known competing financial interests or personal relationships that could have appeared to ifluence the work reported in this paper.

\section{Data availability}

Data will be made available on request to the corresponding authors.

\section{Author contributions}
All authors contributed to the experimental preparation and on-line  data collection.
Chengbo Li contributed to the conceptualization, methodology and design. 
Data analysis was performed by Chengbo Li, Qungang Wen, Huiming Jia. 
The first draft of the manuscript was written by Chengbo Li, and all authors commented on previous versions of the manuscript. All authors read and approved the final manuscript.

\printcredits



\bibliography{myrefs}

@preamble{{\providecommand{\noopsort}[1]{} \providecommand{\singleletter}[1]{#1}%}}

@article{RMP2014,
 author = {Back, B. B. and Esbensen, H. and Jiang, C. L. and et al.},
 doi = {10.1103/RevModPhys.86.317},
 issue = {1},
 journal = {Rev. Mod. Phys.},
 month = {Mar},
 numpages = {44},
 pages = {317--360},
 publisher = {American Physical Society},
 title = {Recent developments in heavy-ion fusion reactions},
 url = {https://link.aps.org/doi/10.1103/RevModPhys.86.317},
 volume = {86},
 year = {2014}
}

@article{RMP2002,
 author = {Woosley, S. E. and Heger, A. and Weaver, T. A.},
 doi = {10.1103/RevModPhys.74.1015},
 issue = {4},
 journal = {Rev. Mod. Phys.},
 month = {Nov},
 numpages = {0},
 pages = {1015--1071},
 publisher = {American Physical Society},
 title = {The evolution and explosion of massive stars},
 url = {https://link.aps.org/doi/10.1103/RevModPhys.74.1015},
 volume = {74},
 year = {2002}
}

@article{Ast2001,
 author = {Cumming, Andrew and Bildsten, Lars},
 doi = {10.1086/323937},
 journal = {The Astrophysical Journal},
 month = {sep},
 number = {2},
 pages = {L127},
 publisher = {},
 title = {{Carbon Flashes in the Heavy-Element Ocean on Accreting Neutron Stars}},
 url = {https://dx.doi.org/10.1086/323937},
 volume = {559},
 year = {2001}
}

@article{Alm1960,
 author = {Almqvist, E. and Bromley, D. A. and Kuehner, J. A.},
 doi = {10.1103/PhysRevLett.4.515},
 issn = {0031-9007},
 journal = {Physical Review Letters},
 month = {may},
 number = {10},
 pages = {515--517},
 title = {{Resonances in C12 on Carbon Reactions}},
 url = {https://link.aps.org/doi/10.1103/PhysRevLett.4.515},
 volume = {4},
 year = {1960}
}

@article{Bro1960,
 author = {Bromley, D. A. and Kuehner, J. A. and Almqvist, E.},
 doi = {10.1103/PhysRevLett.4.365},
 issn = {0031-9007},
 journal = {Physical Review Letters},
 month = {apr},
 number = {7},
 pages = {365--367},
 title = {{Resonant Elastic Scattering of C12 by Carbon}},
 url = {https://link.aps.org/doi/10.1103/PhysRevLett.4.365},
 volume = {4},
 year = {1960}
}

@article{Pat1969,
 author = {Patterson, J. R. and Winkler, H. and Zaidins, C. S.},
 doi = {10.1086/150073},
 issn = {0004-637X},
 journal = {The Astrophysical Journal},
 month = {jul},
 number = {9},
 pages = {367},
 title = {{Experimental Investigation of the Stellar Nuclear Reaction $\rm {}^{12}C+{}^{12}C$ at Low Energies}},
 url = {http://adsabs.harvard.edu/doi/10.1086/150073},
 volume = {157},
 year = {1969}
}

@article{Maz1973,
 author = {Mazarakis, Michael G. and Stephens, William E.},
 doi = {10.1103/PhysRevC.7.1280},
 issn = {05562813},
 journal = {Physical Review C},
 number = {4},
 pages = {1280--1287},
 title = {{Experimental measurements of the $\rm {}^{12}C+{}^{12}C$ nuclear reactions at low energies}},
 volume = {7},
 year = {1973}
}

@article{Hig1977,
 author = {High, M.D. and {\v C}ujec, B.},
 doi = {10.1016/0375-9474(77)90179-8},
 issn = {0375-9474},
 journal = {Nuclear Physics A},
 number = {1},
 pages = {181--188},
 title = {{The $\rm {}^{12}C+{}^{12}C$ sub-coulomb fusion cross section}},
 url = {https://www.sciencedirect.com/science/article/pii/0375947477901798},
 volume = {282},
 year = {1977}
}

@article{Ket1980,
 author = {Kettner, K. U. and Lorenz-Wirzba, H. and Rolfs, C.},
 doi = {10.1007/BF01416030},
 issn = {0340-2193},
 journal = {Z. Phys. A Atoms and Nuclei},
 month = {mar},
 number = {1},
 pages = {65--75},
 title = {{The $\rm {}^{12}C+{}^{12}C$ reaction at subcoulomb energies (I)}},
 url = {http://link.springer.com/10.1007/BF01416030},
 volume = {298},
 year = {1980}
}

@article{Bec1981,
 author = {Becker, H W and Kettner, K U and Rolfs, C and et al.},
 doi = {10.1007/BF01421528},
 issn = {0340-2193},
 journal = {Z. Phys. A Atoms and Nuclei},
 month = {dec},
 number = {4},
 pages = {305--312},
 title = {{The $\rm {}^{12}C+{}^{12}C$ reaction at subcoulomb energies (II)}},
 url = {http://link.springer.com/10.1007/BF01421528},
 volume = {303},
 year = {1981}
}

@article{Das1982,
 author = {Dasmahapatra, Binay and {\v C}ujec, Bibiana and Lahlou, Fouad},
 doi = {10.1016/0375-9474(82)90316-5},
 issn = {0375-9474},
 journal = {Nuclear Physics A},
 number = {1},
 pages = {257--272},
 title = {{Fusion Cross Sections for $\rm {}^{12}C+{}^{12}C$, $\rm {}^{12}C+{}^{13}C$ and $\rm {}^{13}C+{}^{13}C$ at Low Energies}},
 volume = {384},
 year = {1982}
}

@article{Agu2006,
 author = {Aguilera, E. F. and Rosales, P. and Martinez-Quiroz, E. and et al.},
 doi = {10.1103/PhysRevC.73.064601},
 issn = {05562813},
 journal = {Physical Review C - Nuclear Physics},
 number = {6},
 pages = {1--12},
 title = {{New $\gamma$-ray measurements for $\rm {}^{12}C+{}^{12}C$ sub-Coulomb fusion: Toward data unification}},
 volume = {73},
 year = {2006}
}

@article{Bar2006,
 author = {Barr{\'o}n-Palos, L. and Aguilera, E.F. and Aspiazu, J. and et al.},
 doi = {10.1016/j.nuclphysa.2006.09.004},
 issn = {0375-9474},
 journal = {Nuclear Physics A},
 pages = {318--332},
 title = {{Absolute cross sections measurement for the $\rm {}^{12}C+{}^{12}C$ system at astrophysically relevant energies}},
 url = {https://www.sciencedirect.com/science/article/pii/S0375947406006282},
 volume = {779},
 year = {2006}
}

@article{Spi2007,
 author = {Spillane, T. and Raiola, F. and Rolfs, C. and et al.},
 doi = {10.1103/PhysRevLett.98.122501},
 issn = {10797114},
 journal = {Physical Review Letters},
 month = {mar},
 number = {12},
 pages = {122501},
 title = {{$\rm {}^{12}C+{}^{12}C$ fusion reactions near the Gamow energy}},
 url = {https://link.aps.org/doi/10.1103/PhysRevLett.98.122501},
 volume = {98},
 year = {2007}
}

@article{Zic2018,
 author = {Zickefoose, J. and {Di Leva}, A. and Strieder, F. and et al.},
 doi = {10.1103/PhysRevC.97.065806},
 issue = {6},
 journal = {Phys. Rev. C},
 month = {Jun},
 numpages = {9},
 pages = {065806},
 publisher = {American Physical Society},
 title = {{Measurement of the $^{12}\mathrm{C}(^{12}\mathrm{C},p)^{23}\mathrm{Na}$ cross section near the Gamow energy}},
 url = {https://link.aps.org/doi/10.1103/PhysRevC.97.065806},
 volume = {97},
 year = {2018}
}

@article{Jcl2018,
 author = {Jiang, C. L. and Santiago-Gonzalez, D. and Almaraz-Calderon, S. and et al.},
 doi = {10.1103/PhysRevC.97.012801},
 issn = {24699993},
 journal = {Physical Review C},
 number = {1},
 pages = {2--7},
 publisher = {American Physical Society},
 title = {{Reaction rate for carbon burning in massive stars}},
 volume = {97},
 year = {2018}
}

@article{Txd2019,
 author = {Tang, Xiao Dong and Ma, Shao Bo and Fang, Xiao and et al.},
 doi = {10.1007/s41365-019-0652-9},
 isbn = {0123456789},
 issn = {22103147},
 journal = {Nuclear Science and Techniques},
 number = {8},
 title = {{An efficient method for mapping the $\rm {}^{12}C+{}^{12}C$ molecular resonances at low energies}},
 volume = {30},
 year = {2019}
}

@article{Bec2020,
 author = {Beck, C. and Mukhamedzhanov, A. M. and Tang, X.},
 doi = {10.1140/epja/s10050-020-00075-2},
 issn = {1434601X},
 journal = {European Physical Journal A},
 number = {3},
 pages = {1--5},
 publisher = {Springer Berlin Heidelberg},
 title = {{Status on $\rm {}^{12}C+{}^{12}C$ fusion at deep subbarrier energies: impact of resonances on astrophysical $\rm S^*$-factors}},
 url = {https://doi.org/10.1140/epja/s10050-020-00075-2},
 volume = {56},
 year = {2020}
}

@article{Fru2020,
 author = {Fruet, G. and Courtin, S. and Heine, M. and et al.},
 doi = {10.1103/PhysRevLett.124.192701},
 issn = {10797114},
 journal = {Physical Review Letters},
 number = {19},
 pages = {192701},
 pmid = {32469543},
 publisher = {American Physical Society},
 title = {{Advances in the Direct Study of Carbon Burning in Massive Stars}},
 url = {https://doi.org/10.1103/PhysRevLett.124.192701},
 volume = {124},
 year = {2020}
}

@article{Lyj2020,
 author = {Li, Y. J. and Fang, X. and Bucher, B. and et al.},
 doi = {10.1088/1674-1137/abae56},
 issn = {16741137},
 journal = {Chinese Physics C},
 month = {nov},
 number = {11},
 pages = {115001},
 title = {{Modified astrophysical S-factor of $\rm {}^{12}C+{}^{12}C$ fusion reaction at sub-barrier energies}},
 url = {https://iopscience.iop.org/article/10.1088/1674-1137/abae56},
 volume = {44},
 year = {2020}
}

@article{Tan2020,
 author = {Tan, W. P. and Boeltzig, A. and Dulal, C. and et al.},
 doi = {10.1103/PhysRevLett.124.192702},
 issn = {0031-9007},
 journal = {Physical Review Letters},
 month = {may},
 number = {19},
 pages = {192702},
 pmid = {32469557},
 publisher = {American Physical Society},
 title = {{New Measurement of $\rm {}^{12}C+{}^{12}C$ Fusion Reaction at Astrophysical Energies}},
 url = {https://doi.org/10.1103/PhysRevLett.124.192702 https://link.aps.org/doi/10.1103/PhysRevLett.124.192702},
 volume = {124},
 year = {2020}
}

@article{Znt2020,
 author = {Zhang, N.T. and Wang, X.Y. and Tudor, D. and et al.},
 doi = {10.1016/j.physletb.2019.135170},
 issn = {03702693},
 journal = {Physics Letters B},
 month = {feb},
 pages = {135170},
 title = {{Constraining the $\rm {}^{12}C+{}^{12}C$ astrophysical S-factors with the $\rm {}^{12}C+{}^{13}C$ measurements at very low energies}},
 url = {https://linkinghub.elsevier.com/retrieve/pii/S0370269319308925},
 volume = {801},
 year = {2020}
}

@article{Muk2022,
 author = {Mukhamedzanov, A. M.},
 doi = {10.1140/epja/s10050-022-00718-6},
 issn = {1434601X},
 journal = {European Physical Journal A},
 number = {4},
 pages = {1--12},
 publisher = {Springer Berlin Heidelberg},
 title = {{Status of deep subbarrier $\rm {}^{12}C+{}^{12}C$ fusion and advancing the Trojan horse method}},
 url = {https://doi.org/10.1140/epja/s10050-022-00718-6},
 volume = {58},
 year = {2022}
}

@article{tan2024,
  title = {Coincident measurement of the $^{12}\mathrm{C}+^{12}\mathrm{C}$ fusion cross section via the differential thick-target technique},
  author = {Tan, W. P. and Gula, A. and Lee, K. and Majumdar, A. and Moylan, S. and Olivas-Gomez, O. and Shahina and Wiescher, M. and Aguilera, E. F. and Lizcano, D. and Martinez-Quiroz, E. and Morales-Rivera, J. C.},
  journal = {Phys. Rev. C},
  volume = {110},
  issue = {3},
  pages = {035808},
  numpages = {9},
  year = {2024},
  month = {Sep},
  publisher = {American Physical Society},
  doi = {10.1103/PhysRevC.110.035808},
  url = {https://link.aps.org/doi/10.1103/PhysRevC.110.035808}
}

@article{nip2025,
  title = {Refining the deep sub-barrier $^{12}\mathrm{C}+^{12}\mathrm{C}$ fusion excitation function with the STELLA apparatus},
  author = {Nippert, J. and Courtin, S. and Heine, M. and Jenkins, D. G. and Adsley, P. and Bonhomme, A. and Canavan, R. and Curien, D. and Dumont, T. and Gregor, E. and Harmant, G. and Monpribat, E. and Morrison, L. and Moukaddam, M. and Richer, M. and Rudigier, M. and Romero, J. G. Vega and Catford, W. N. and Cotte, P. and Della Negra, S. and Haefner, G. and Hammache, F. and Lesrel, J. and Pascu, S. and Podoly\'ak, Zs. and Regan, P. H. and Ribaud, I. and de S\'er\'eville, N. and Stodel, C. and Vesi\ifmmode \acute{c}\else \'{c}\fi{}, J.},
  collaboration = {STELLA Collaboration},
  journal = {Phys. Rev. C},
  volume = {111},
  issue = {6},
  pages = {065804},
  numpages = {7},
  year = {2025},
  month = {Jun},
  publisher = {American Physical Society},
  doi = {10.1103/PhysRevC.111.065804},
  url = {https://link.aps.org/doi/10.1103/PhysRevC.111.065804}
}

@article{wyb2025,
 author = {Nan, Weike and Wang, Youbao and Su, Jun and et al.},
 doi = {10.1016/j.physletb.2025.139341},
 issn = {0370-2693},
 journal = {Physics Letters B},
 langid = {english},
 pages = {139341},
 title = {Determination of Resonances in {{Gamow}} Window for {$\rm {}^{12}C+{}^{12}C$} Fusion Reaction via Thick-Target Inverse Kinematics Method},
 volume = {862},
 year = {2025}
}

@article{Tum2018,
 author = {Tumino, A. and Spitaleri, C. and {La Cognata}, M. and et al.},
 doi = {10.1038/s41586-018-0149-4},
 issn = {0028-0836},
 journal = {Nature},
 month = {may},
 number = {7707},
 pages = {687--690},
 title = {{An increase in the $\rm {}^{12}C+{}^{12}C$ fusion rate from resonances at astrophysical energies}},
 url = {http://www.nature.com/articles/s41586-018-0149-4},
 volume = {557},
 year = {2018}
}

@article{Tum2018b,
 archiveprefix = {arXiv},
 arxivid = {1807.06148},
 author = {Tumino, A. and Spitaleri, C. and {La Cognata}, M. and et al.},
 eprint = {1807.06148},
 month = {jul},
 number = {July},
 title = {{Reply to the Comments on the $\rm {}^{12}C+{}^{12}C$ fusion $S^*$-factor}},
 url = {http://arxiv.org/abs/1807.06148},
 year = {2018}
}

@article{Muk2018,
 archiveprefix = {arXiv},
 arxivid = {1806.05921},
 author = {Mukhamedzhanov, Akram and Tang, Xiaodong and Pang, Danyang},
 eprint = {1806.05921},
 month = {jun},
 pages = {5--6},
 title = {{Comments on the $\rm {}^{12}C+{}^{12}C$ fusion $S^*$-factor}},
 url = {http://arxiv.org/abs/1806.05921},
 year = {2018}
}

@article{Muk2019,
 author = {Mukhamedzhanov, A. M. and Pang, D. Y. and Kadyrov, A. S.},
 doi = {10.1103/PhysRevC.99.064618},
 issn = {24699993},
 journal = {Physical Review C},
 number = {6},
 pages = {1--12},
 title = {{Astrophysical factors of $\rm {}^{12}C+{}^{12}C$ fusion extracted using the Trojan horse method}},
 volume = {99},
 year = {2019}
}

@article{Bon2020,
  title = {Calculation of the $^{12}\mathrm{C}+^{12}\mathrm{C}$ sub-barrier fusion cross section in an imaginary-time-dependent mean field theory},
  author = {Bonasera, A. and Natowitz, J. B.},
  journal = {Phys. Rev. C},
  volume = {102},
  issue = {6},
  pages = {061602},
  numpages = {4},
  year = {2020},
  month = {Dec},
  publisher = {American Physical Society},
  doi = {10.1103/PhysRevC.102.061602},
  url = {https://link.aps.org/doi/10.1103/PhysRevC.102.061602}
}

@article{Bau1986,
 author = {Baur, G.},
 doi = {10.1016/0370-2693(86)91483-8},
 issn = {03702693},
 journal = {Physics Letters B},
 month = oct,
 number = {2-3},
 pages = {135--138},
 title = {{Breakup Reactions as an Indirect Method to Investigate Low-Energy Charged-Particle Reactions Relevant for Nuclear Astrophysics}},
 volume = {178},
 year = {1986}
}

@article{Typ2000,
 author = {Typel, S. and Wolter, H. H.},
 doi = {10.1007/s006010070010},
 issn = {01777963},
 journal = {Few-Body Systems},
 month = nov,
 number = {1},
 pages = {75--93},
 title = {{Extraction of Astrophysical Cross Sectionsin the Trojan-Horse Method}},
 volume = {29},
 year = {2000}
}

@article{Typ2003,
 author = {Typel, S. and Baur, G.},
 doi = {10.1016/S0003-4916(03)00060-5},
 issn = {00034916},
 journal = {Annals of Physics},
 month = jun,
 number = {2},
 pages = {228--265},
 title = {{Theory of the Trojan-Horse Method}},
 volume = {305},
 year = {2003}
}

@article{Bau2004,
 author = {Baur, Gerhard and Typel, Stefan},
 doi = {10.1143/PTPS.154.333},
 issn = {0375-9687},
 journal = {Progress of Theoretical Physics Supplement},
 pages = {333--340},
 title = {{Theory of the Trojan-Horse Method}},
 volume = {154},
 year = {2004}
}

@article{Tri2014,
 author = {Tribble, R. E. and Bertulani, C. A. and Cognata, M. La and et al.},
 doi = {10.1088/0034-4885/77/10/106901},
 issn = {0034-4885},
 journal = {Reports on Progress in Physics},
 month = oct,
 number = {10},
 pages = {106901},
 title = {{Indirect Techniques in Nuclear Astrophysics: A Review}},
 volume = {77},
 year = {2014}
}

@article{Ber2018,
 author = {Bertulani, C.A. and Hussein, M.S. and Typel, S.},
 doi = {10.1016/j.physletb.2017.11.050},
 issn = {03702693},
 journal = {Physics Letters B},
 month = jan,
 pages = {217--221},
 publisher = {Elsevier B.V.},
 title = {Assessing the Foundation of the {{Trojan Horse Method}}},
 volume = {776},
 year = {2018}
}

@article{epjrev2019,
 author = {Spitaleri, C. and {La Cognata}, M. and Lamia, L. and et al.},
 doi = {10.1140/epja/i2019-12833-0},
 isbn = {2019128330},
 issn = {1434-6001},
 journal = {The European Physical Journal A},
 month = sep,
 number = {9},
 pages = {161},
 title = {Astrophysics Studies with the {{Trojan Horse Method}}},
 volume = {55},
 year = {2019}
}

@article{prl2007,
 author = {Tumino, A. and Spitaleri, C. and Mukhamedzhanov, A. and et al.},
 doi = {10.1103/PhysRevLett.98.252502},
 issn = {0031-9007},
 journal = {Physical Review Letters},
 month = jun,
 number = {25},
 pages = {252502},
 title = {Suppression of the {{Coulomb Interaction}} in the {{Off-Energy-Shell}} p-p {{Scattering}} from the p+d\ensuremath{\rightarrow}p+p+n {{Reaction}}},
 volume = {98},
 year = {2007}
}

@article{lac2007,
  title = {Astrophysical {S(E)} factor of the $^{15}\mathrm{N}$($p,\ensuremath{\alpha}$)$^{12}\mathrm{C}$ reaction at sub-Coulomb energies via the Trojan horse method},
  author = {La Cognata, M. and Romano, S. and Spitaleri, C. and Cherubini, S. and Crucill\`a, V. and Gulino, M. and Lamia, L. and Pizzone, R. G. and Tumino, A. and Tribble, R. and Fu, Changbo and Goldberg, V. Z. and Mukhamedzhanov, A. M. and Schmidt, D. and Tabacaru, G. and Trache, L. and Irgaziev, B. F.},
  journal = {Phys. Rev. C},
  volume = {76},
  issue = {6},
  pages = {065804},
  numpages = {15},
  year = {2007},
  month = {Dec},
  publisher = {American Physical Society},
  doi = {10.1103/PhysRevC.76.065804},
  url = {https://link.aps.org/doi/10.1103/PhysRevC.76.065804}
}

@article{prl2012,
 author = {{La Cognata}, M. and Spitaleri, C. and Trippella, O. and et al.},
 doi = {10.1103/PhysRevLett.109.232701},
 issn = {0031-9007},
 journal = {Physical Review Letters},
 month = dec,
 number = {23},
 pages = {232701},
 pmid = {23368189},
 title = {{Measurement of the -3 keV Resonance in the Reaction 13C(a,n)16O of Importance in the s-Process}},
 volume = {109},
 year = {2012}
}

@article{plb2015,
 author = {Lombardo, I. and Dell'Aquila, D. and {Di Leva}, A. and et al.},
 doi = {10.1016/j.physletb.2015.06.073},
 issn = {03702693},
 journal = {Physics Letters B},
 month = sep,
 pages = {178--182},
 publisher = {Elsevier B.V.},
 title = {{Toward a Reassessment of the $^{19}F(p,\alpha_0)^{16}O$ Reaction Rate at Astrophysical Temperatures}},
 volume = {748},
 year = {2015}
}

@article{apj2017,
 author = {Pizzone, R. G. and D.Agata, G. and Cognata, M. La and et al.},
 doi = {10.3847/1538-4357/836/1/57},
 issn = {1538-4357},
 journal = {The Astrophysical Journal},
 lccn = {Not Found},
 month = feb,
 number = {1},
 pages = {57},
 title = {{First Measurement of the $^{19}F(\alpha,p)^{22}Ne$ Reaction at Energies of Astrophysical Relevance}},
 volume = {836},
 year = {2017}
}

@article{wen2008,
 author = {Wen, Qun-Gang and Li, Cheng-Bo and Zhou, Shu-Hua and et al.},
 doi = {10.1103/PhysRevC.78.035805},
 issn = {0556-2813},
 journal = {Physical Review C},
 langid = {american},
 month = sep,
 number = {3},
 pages = {035805},
 title = {{Trojan Horse Method Applied to $^9Be(p,\alpha)^6Li$ at Astrophysical Energies}},
 volume = {78},
 year = {2008}
}

@article{wen2011,
 author = {Wen, Qun-Gang and Li, Cheng-Bo and Zhou, Shu-Hua and et al.},
 doi = {10.1088/0954-3899/38/8/085103},
 issn = {0954-3899, 1361-6471},
 journal = {Journal of Physics G: Nuclear and Particle Physics},
 langid = {english},
 month = aug,
 number = {8},
 pages = {085103},
 title = {{A New Approach to Select the Quasifree Mechanism in the Trojan Horse Method}},
 urldate = {2025-01-16},
 volume = {38},
 year = {2011}
}

@article{lcb2015a,
 author = {Li, Cheng-Bo and Wen, Qun-Gang and Zhou, Shu-Hua and et al.},
 doi = {10.1088/1674-1137/39/5/054001},
 issn = {1674-1137},
 journal = {Chinese Physics C},
 month = may,
 number = {5},
 pages = {054001},
 title = {{Experimental Spectra Analysis in THM with the Help of Simulation Based on the Geant4 Framework}},
 volume = {39},
 year = {2015}
}

@article{lcb2015,
 author = {Li, Chengbo and Wen, Qungang and Fu, Yuanyong and et al.},
 copyright = {http://link.aps.org/licenses/aps-default-license},
 doi = {10.1103/PhysRevC.92.025805},
 issn = {0556-2813, 1089-490X},
 journal = {Physical Review C},
 langid = {english},
 month = aug,
 number = {2},
 pages = {025805},
 title = {{Measurement of the $^2H(d,p)^3H$ Reaction at Astrophysical Energies via the Trojan-Horse Method}},
 urldate = {2025-01-16},
 volume = {92},
 year = {2015}
}

@article{wen2016,
 author = {Wen, Qun-Gang and Li, Cheng-Bo and Zhou, Shu-Hua and et al.},
 copyright = {http://link.aps.org/licenses/aps-default-license},
 doi = {10.1103/PhysRevC.93.035803},
 issn = {2469-9985, 2469-9993},
 journal = {Physical Review C},
 langid = {english},
 month = mar,
 number = {3},
 pages = {035803},
 title = {{Experimental Study to Explore the $^8Be$-Induced Nuclear Reaction via the Trojan Horse Method}},
 urldate = {2025-01-16},
 volume = {93},
 year = {2016}
}

@article{lcb2017,
 author = {Li, Chengbo and Wen, Qungang and Tumino, A. and et al.},
 copyright = {http://link.aps.org/licenses/aps-default-license},
 doi = {10.1103/PhysRevC.95.035804},
 issn = {2469-9985, 2469-9993},
 journal = {Physical Review C},
 langid = {english},
 month = mar,
 number = {3},
 pages = {035804},
 title = {{Beam-Energy Dependence and Updated Test of the Trojan-Horse Nucleus Invariance via a Measurement of the $^2H(d,p)^3H$ Reaction at Low Energies}},
 urldate = {2025-01-16},
 volume = {95},
 year = {2017}
}

@article{wang2024,
 author = {Wang, Xue-Jian and Wen, Qun-Gang and Li, Cheng-Bo and et al.},
 doi = {10.1016/j.physletb.2024.138745},
 issn = {03702693},
 journal = {Physics Letters B},
 langid = {english},
 month = jul,
 pages = {138745},
 title = {{Studying Subthreshold Resonance Using the Trojan Horse Method}},
 urldate = {2025-01-15},
 volume = {854},
 year = {2024}
}

@article{CF1988,
 author = {Caughlan, Georgeanne R. and Fowler, William A.},
 doi = {10.1016/0092-640x(88)90009-5},
 issn = {0092-640X},
 journal = {Atomic Data and Nuclear Data Tables},
 number = {2},
 pages = {283--334},
 title = {Thermonuclear Reaction Rates {{V}}},
 volume = {40},
 year = {1988}
}

@article{Hin2007,
 author = {Jiang, C. L. and Rehm, K. E. and Back, B. B. and et al.},
 doi = {10.1103/PhysRevC.75.015803},
 issue = {1},
 journal = {Phys. Rev. C},
 month = {Jan},
 numpages = {11},
 pages = {015803},
 publisher = {American Physical Society},
 title = {Expectations for $^{12}\mathrm{C}$ and $^{16}\mathrm{O}$ induced fusion cross sections at energies of astrophysical interest},
 url = {https://link.aps.org/doi/10.1103/PhysRevC.75.015803},
 volume = {75},
 year = {2007}
}

@article{Cop2009,
 author = {Cooper, Randall L and Steiner, Andrew W and Brown, Edward F},
 doi = {10.1088/0004-637X/702/1/660},
 journal = {Astrophysical Journal},
 number = {1},
 title = {Possible Resonances in the {$\rm {}^{12}C+{}^{12}C$} Fusion Rate and Superburst Ignition},
 volume = {702},
 year = {2009}
}

@article{AMD2021,
 author = {Taniguchi, Yasutaka and Kimura, Masaaki},
 doi = {10.1016/j.physletb.2021.136790},
 issn = {0370-2693},
 journal = {Physics Letters B},
 pages = {136790},
 title = {{$\rm {}^{12}C+{}^{12}C$ Fusion S-Factor from a Full-Microscopic Nuclear Model}},
 volume = {823},
 year = {2021}
}

@article{DIM2024,
 author = {Taniguchi, Yasutaka and Kimura, Masaaki},
 doi = {10.1016/j.physletb.2023.138434},
 issn = {0370-2693},
 journal = {Physics Letters B},
 pages = {138434},
 title = {{Impact of the Molecular Resonances on the {$\rm {}^{12}C+{}^{12}C$} Fusion Reaction Rate}},
 volume = {849},
 year = {2024}
}

@article{a16O1984,
 author = {Bauhoff, W. and Schultheis, H. and Schultheis, R.},
 doi = {10.1103/PhysRevC.29.1046},
 issue = {3},
 journal = {Phys. Rev. C},
 month = {Mar},
 numpages = {0},
 pages = {1046--1055},
 publisher = {American Physical Society},
 title = {Alpha cluster model and the spectrum of $^{16}\mathrm{O}$},
 url = {https://link.aps.org/doi/10.1103/PhysRevC.29.1046},
 volume = {29},
 year = {1984}
}

@article{a16O1988,
 author = {Cseh, J. and L{\'e}vai, G.},
 doi = {10.1103/PhysRevC.38.972},
 issue = {2},
 journal = {Phys. Rev. C},
 month = {Aug},
 numpages = {0},
 pages = {972--976},
 publisher = {American Physical Society},
 title = {Core-plus-alpha-particle states of $^{20}\mathrm{Ne}$ and $^{16}\mathrm{O}$ in terms of vibron models},
 url = {https://link.aps.org/doi/10.1103/PhysRevC.38.972},
 volume = {38},
 year = {1988}
}

@article{APJ1996,
  title={Indirect Investigation of the d + 6Li Reaction at Low Energies Relevant for Nuclear Astrophysics},
  author={ Cherubini, S  and  Kondratyev, V. N  and  Lattuada, M  and  Spitaleri, C  and  Miljanic, D  and  Zadro, M  and  Baur, G },
  journal={Astrophysical Journal},
  volume={457},
  number={2},
  pages={855},
  year={1996},
}

@article{PRC2001,
  title={Trojan horse method applied to $^2H(^6Li,\alpha)^4He$ at astrophysical energies},
  author={ Spitaleri, C.  and  Typel, S.  and  Pizzone, R. G.  and  Aliotta, M.  and  Zadro, M. },
  journal={Physical Review C},
  volume={63},
  year={2001},
}

@article{EPJA2020,
  title={Study of the quasi-free $^3He + ^9Be \rightarrow 3 \alpha$  reaction for the Trojan Horse Method },
  author={ Spitaleri, C.  and  Lattuada, M.  and  Cvetinovic, A.  and  et al.},
  journal={The European Physical Journal A},
  volume={56},
  year={2020},
  doi = {10.1140/epja/s10050-020-00026-x},
}

@article{EPJA2021,
  title={The $^3He + ^5He \rightarrow \alpha + \alpha$ reaction below the Coulomb barrier via the Trojan Horse Method},
  author={ Spitaleri, C.  and  Typel, S.  and  Bertulani, C.  and  Mukhamedzhanov, A.  and  Kajino, T.  and  Lattuada, M.  and  Cvetinovic, A.  and  Messina, S.  and  Guardo, G.  and  Soi, N. },
  journal={The European Physical Journal A},
  volume={57},
  year={2021},
  doi = {10.1140/epja/s10050-020-00324-4},
}




\end{document}